# Long-term monitoring of WASP-19 b: signs of apsidal precession and molecular signatures

A. R. Rajkumar[1,2], A. Bayo[2], P. Peng[3,4], J. Tregloan-Reed[5], J. Southworth[6], T. C. Hinse[7], L. G. Alegre[8,9], F. Amadio[10], M. Andersen[10], N. Bach-Moller[10], M. Basilicata[11], M. Bonavita[9], V. Bozza[12,13], M. J. Burgdorf[14], R. E. Cannon[9], G. Columba[12], M. Dominik[15], A. Donaldson[9], R. Figuera Jaimes[1,15,16,17], J. Fynbo[10], M. Hundertmark[18], U. G. Jorgensen[10], E. Khalouei[19], H. Korhonen[20,21], P. Longa-Pena[5], M. Rabus[22], S. Rahvar[22], H. Rendell-Bhatti[9], P. Rota[12], A. Rozek[9], S. Sajadian[23], J. Skottfelt[24], C. Snodgrass[9]

*(Affiliations can be found after the references)*

**ABSTRACT**

*Context.* With over 6,000 exoplanets discovered to date, approximately 12 % are classified as hot-Jupiters. Due to their large sizes and short orbital periods ($P < 10$ day), they are easier to detect and provide crucial insights into planetary formation, atmospheric properties, and orbital dynamics. Among these, ultra-short-period exoplanets ($P \leq 1$ d) are particularly interesting, as they are expected to undergo orbital decay driven by strong tidal interactions. Despite theoretical predictions, WASP-12 b and WASP-4 b remain the confirmed hot-Jupiters experiencing measurable orbital decay.

*Aims.* This study presents a homogeneous analysis of WASP-19 b to investigate both its orbital dynamics and atmospheric composition. Leveraging a 15-year dataset, our goal is to assess whether the system exhibits long-term deviations from a constant orbital period and to investigate whether any detected variations are consistent with tidal orbital decay, apsidal precession, or periodic signals indicative of a potential planetary perturber. Additionally, we also construct a photometric transmission spectrum to characterize its atmosphere.

*Methods.* We analyze multi-wavelength light curves, incorporating starspot modeling with `PRISM` to account for stellar inhomogeneities. To assess orbital evolution, we fit linear, quadratic, and cubic ephemeris models to transit timing residuals with respect to a non-decaying orbit.

*Results.* Our analysis, which includes 27 new transits, reveals no statistically significant periodic signal in the transit timings. Although none of the tested ephemeris models fully reproduce the observed timing scatter, the mid-transit times exhibit systematic deviations from a strictly constant orbital period and are best reproduced by the cubic ephemeris in a relative model-comparison sense, indicating a slow, non-periodic long-term trend over the ~15-year baseline. This behavior is more consistent with gradual orbital precession than with monotonic tidal decay, for which a dominant quadratic trend would be expected. Fitting a precession model yields a rate of $\dot{\omega}_{\mathrm{obs}} = (1.00 \pm 0.12) \times 10^{-4}$ rad/orbit, corresponding to a planetary Love number $k_{2p} = 0.107 \pm 0.08$, in agreement with recent independent estimates. The transmission spectrum reveals signatures of Na, K, and $H_2O$, with no strong evidence of TiO/VO, likely due to the resolution limits of the photometric data.

*Conclusions.* Our results support that apsidal precession could be the dominant process governing the long-term orbital evolution of WASP-19 b, possibly sustained by weak eccentricity forcing from the wide companion WASP-19 B. These orbital dynamics can, in turn, impact the atmospheric structure by modulating the irradiation history, potentially altering molecular abundances over time. Our findings highlight the importance of combining TTV analyzes with multi-wavelength atmospheric data, while emphasizing that additional high-quality timing and spectroscopic observations are required to corroborate the fidelity of the proposed orbital model.

## 1. Introduction

The discovery of the first hot-Jupiter, 51 Pegasi b, raised numerous unresolved questions regarding established theories of planetary formation and the mechanisms driving the migration and orbital decay of short-period gas giant planets (Mayor & Queloz 1995). Planetary system formation remains an active area of research, with two widely accepted theories proposing that Jupiter-like planets form either via core accretion (Pollack et al. 1996) and/or through gravitational instability (Boss 1997), typically at significant distances from the host star. In the core accretion (CA) model, dust particles in the protoplanetary disk coagulate into planetesimals, which grow into solid planetary cores; once these reach a critical mass of $\sim 10\,M_{\oplus}$, they begin accreting gas from the disk (Mordasini 2018). However, the short lifetime of protoplanetary disks (~1–10 Myr; Haisch et al. 2001) presents a major challenge to this model, particularly for the timely formation of gas giants. This has led to the introduction of the pebble accretion mechanism, which significantly accelerates core growth by allowing centimeter- to meter-sized particles to be efficiently captured by planetary embryos (Lambrechts & Johansen 2014; Bitsch et al. 2015). Alternatively, the gravitational instability (GI) model suggests that in massive and rapidly cooling disks, direct fragmentation into self-gravitating clumps can occur (Boss 1997); however, GI is often limited by its requirement for very high disk masses and efficient cooling, making it less favored for forming close-in planets. These regions in the disk, rich in condensed material, provide favorable conditions for the onset of such formation processes. Subsequently, the resulting planets migrate inward to close-in orbits (< 0.05 au) through a variety of mechanisms, such as planet-planet scattering (Chatterjee et al. 2008), high eccentricity migration (Rasio & Ford 1996), or interactions with the gas disk itself (Paardekooper et al. 2023).

Once these planets reach extremely close-in orbits, they are subjected to intense tidal forces from their host stars, which lead to gradual orbital evolution. The physical mechanism of tidal decay in hot-Jupiter systems involves the transfer of energy and angular momentum between the planet and its host star due to the mutual gravitational interaction (Hebb et al. 2010; Patra et al. 2017; Maciejewski et al. 2018; Mathis 2019). As hot-Jupiters or-bit very close to their host star (P < 10 d), the gravitational pull is stronger on the side of the planet facing the star compared to the side facing away from it. This variation in gravitational force leads to a distortion of the planet's shape. As the planet undergoes this deformation, it tries to adapt to the gravitational pull by reshaping itself. However, this continuous reshaping of the planet results in the generation of internal heat within the planet through tidal dissipation (e.g., Gu et al. 2003; Ogilvie & Lin 2004). The heat generated by tidal dissipation represents a conversion of orbital energy into thermal energy within the planet. This process can contribute to the inflation of the gaseous envelope; however, current inflation models do not fully reproduce the observed radii of all hot-Jupiters, implying that tidal heating is only one of several mechanisms that may operate with varying efficiency across the population. Specifically, when the internal heat increases, it raises the temperature within the planet, reducing the density of the outer layers and causing them to swell. This process of envelope inflation is particularly relevant for hot and ultra-hot Jupiters, where the added thermal energy can significantly alter the planet's radius and overall structure (Hou & Wei 2022). Moreover, this process affects the planet's transmission spectrum by modifying vertical temperature gradients and driving atmospheric circulation and influences the abundances of molecular species (Komacek & Showman 2016).

Tidal forces affect the planet's rotation (Barker et al. 2016), leading to energy loss from its orbital motion and causing its orbit to shrink—a phenomenon known as orbital decay (Ogilvie 2014). This process leads to a shortening of the planet's orbital period, causing transits to occur earlier than predicted. As the planet's semi-major axis decreases, it spirals closer to its host star over time. The energy loss also influences the planet's rotation, driving it toward an aligned spin with its orbit (Gallet et al. 2017), and so becoming tidally locked with its host star. This stellar irradiation drives photochemical processes in the atmosphere of the hot-Jupiters, potentially altering the molecular abundances detectable in the transmission spectrum.

Predicting the long-term orbital evolution of hot-Jupiters remains challenging due to the complex interplay between stellar and planetary properties that influence tidal dissipation. Although intense stellar tides can gradually shrink planetary orbits, potentially leading to planetary engulfment, additional processes, such as magnetic braking (Dawson 2014, and references therein), can further affect orbital alignment and evolution by slowing stellar rotation through the loss of angular momentum to stellar winds. These coupled interactions make it difficult to model decay rates precisely, especially in systems where unseen companions or evolving stellar interiors may also play a role. However, ultra-short-period systems like WASP-19 b ($P \approx 19$ hr; Hebb et al. 2010) can enhance our understanding and refine our predictive capabilities, offering valuable insights into the ultimate fate of hot-Jupiters.

Moreover, the study of transmission spectra, when combined with transit timing variations (TTVs), offers a comprehensive approach to understanding the interplay between tidal decay, internal heating, and atmospheric dynamics. By analyzing changes in the transmission spectrum over time, researchers can constrain the chemical and physical processes that occur in the atmosphere as a result of tidal interactions and migration. This is especially important for systems like WASP-19, where the strong tidal forces and close proximity to the host star make these effects particularly significant, and they provide a critical diagnostic for investigating the impact of tidal decay and internal heating on atmospheric composition and structure.

It is, therefore, beneficial to look for TTVs in systems with hot-Jupiters. Population studies have provided some evidence that orbital decay does take place on astrophysically significant timescales – for example WASP-12 b Hebb et al. (2010) and WASP-4 b Baştürk et al. (2025) – and the effects of tides are small on human timescales, making it best to focus this search on the most promising systems (Eq. 1, Maciejewski 2019) and (Eq. 5, Birkby et al. 2014). Observations of TTVs can help constrain the efficiency of tidal interactions, track changes in atmospheric features, and improve predictions of the long-term evolution of these systems.

Other mechanisms that can cause a TTV signal include gravitational interactions with additional planets or moons (Ballard et al. 2011), stellar activity (such as starspots or flares; Ioannidis et al. 2016; Sanchis-Ojeda et al. 2011a), variations in the planet's orbital eccentricity and apsidal precession (Antoniciello et al. 2021), Applegate effect (Watson & Marsh 2010), and relativistic effects (Antoniciello et al. 2021).

This paper is structured as follows. In Sect. 2 we describe the hot-Jupiter, WASP-19 b. Section 3 describes the observational datasets and the data reduction method. Sect. 4 focuses on the light curve analysis method. Section 5 explores TTV and apsidal motion analysis. In Sect. 6 we present the transmission spectrum and retrieval results. We summarize our results in Sect. 7 and present a discussion and conclusions in Sects. 8 and 9.

## 2. WASP-19 b

WASP-19 b (Hebb et al. 2010) is a Jupiter-like planet ($M_p$ = 1.11 $M_{Jup}$ & $R_P$ = 1.39 $R_{Jup}$) that orbits an active G8V solar-type star ($M_* = 0.9\ M_\odot$, $R_* = 1.0\ R_\odot$, and an Age ∼ 11 Gyr) and is one of the first ultra-short-period ($P$ = 0.8 days) planets discovered. The host star shows strong magnetic activity, with a chromospheric index of $\log R'_{HK} \approx -4.5$ (Doyle et al. 2014) and a coronal X-ray to bolometric luminosity ratio of $\log(L_X/L_{bol}) \approx -5.1$ (Maggio et al. 2012). In addition, the photometric rotational modulation of ~0.6–0.8% in the optical flux indicates persistent starspots, with a measured stellar rotation period of ∼10.5 days (Hebb et al. 2010; Mancini et al. 2013).

Since its discovery by Hebb et al. (2010), it has been studied in detail and the planet's physical properties have been refined and re-determined by several authors (Cortés-Zuleta et al. 2020; Wong et al. 2016; Sing et al. 2016; Tregloan-Reed et al. 2013; Mancini et al. 2013; Hellier et al. 2011; Lendl et al. 2013). In a system like WASP-19, where the stellar rotation period (10.50 ± 0.20 days Hebb et al. 2010) exceeds the orbital period (0.7888396 d ± 0.0000010 d Hebb et al. 2010), tidal dissipation in the star can transfer angular momentum from the orbit to the stellar spin, causing orbital decay. A few previous studies from Levrard et al. (2009) and Matsumura et al. (2010) predict the significant impact of tidal interactions on the orbital evolution of close-in exoplanets such as WASP-19 b. Calculations from Matsumura et al. (2010) have predicted that the orbit of WASP-19 b may become unstable due to insufficient total angular momentum, possibly leading to tidal disruption. Valsecchi & Rasio (2014) predicted that, due to the convective damping of equilibrium tides, WASP-19 b's transit arrival times would deviate by at least 34 s from a constant-period ephemeris over a 10-year baseline.

The WASP-19 system is especially interesting to study further, as the host star is known to be highly active with prominent starspots Tregloan-Reed et al. (2013). The presence of a convective envelope in the G8V host star can significantly impact the tidal dissipation processes within the system. Additionally, starspots can introduce systematics that affect the accurate measurement of transit midpoints (Sanchis-Ojeda et al. 2011b) and transit depths (Pont et al. 2008). To mitigate these effects, we employ the PRISM (Planetary Retrospective Integrated Starspot Model; Tregloan-Reed et al. 2013, 2015; see our Sect. 4.1) algorithm in our analysis, which allows us to model and account for the impact of starspots on the transit light curves. This allows us to recover the true transit shape and to separate spot-induced anomalies from genuine planetary signals. Through this procedure, we refine the orbital inclination, impact parameter, and mid-transit timing, ensuring that timing residuals are not contaminated by stellar variability. Moreover, by isolating and removing the wavelength-dependent effects of stellar heterogeneity, the derived transmission spectrum becomes more representative of the planetary atmosphere rather than the host-star surface. Consequently, the inclusion of starspot modeling not only improves the precision of the orbital and transit parameters but also enhances the reliability of atmospheric retrievals. This integrated approach strengthens the link between dynamical evolution and the observed transmission properties of WASP-19 b, providing a self-consistent view of the system's long-term behavior.

In this work, we present a comprehensive analysis of 27 new, systematically obtained transit observations of WASP-19 b from the Danish Telescope and more than 100 transits from TESS, combined with available transit light curves from the literature. This extensive dataset allows us to conduct a long-term, detailed investigation of the evolution of the WASP-19 system over the past two decades. By analyzing these transit observations, we aim to test the predictions of the expected orbital decay of WASP-19 b.

## 3. Observations and data reduction

### 3.1. Danish 1.54-m telescope

WASP-19 b data were collected over a 13 year period, from 2015 to 2023, by the MiNDSTEp collaboration [1] using the Danish 1.54-m telescope located at the ESO La Silla Observatory. The observations were performed using the Danish Faint Object Spectrograph and Camera (DFOSC) imager, which has a field of view of $13.7' \times 13.7'$ at a plate scale of $0.39''$ per pixel. However, to decrease the readout time, the charge-coupled device (CCD) was typically windowed to $12' \times 8'$. A total of 27 transit light curves were recorded: 14 in the Johnson-Cousins R filter and 13 in the Johnson-Cousins I filter. Two of these (on 26 April 2016 and 20 May 2022) were incomplete and excluded from the final analysis, leaving 25 full transits considered for detailed investigation. The complete set of light curves is presented in Fig. 1. All scientific images were first calibrated using standard CCD reduction procedures. This included subtracting a master bias frame to correct for electronic readout offsets and applying flat-field corrections to account for pixel-to-pixel sensitivity variations. Although flat-fielding typically produced only marginal improvements, especially when stellar point spread functions (PSFs) remained stable on the detector. This ensured photometric consistency across the field of view.

Photometric extraction was then performed using the defocused photometry of transiting exoplanets (DEFOT) pipeline (Southworth et al. 2009b,a, 2014), which performs aperature photometry through an interactive data language (IDL) implementation of the Dominion Astrophysical Observatory Photometry (DAOPHOT) package (Stetson 1987). The aperture radii were manually selected on a reference frame to minimize the out-of-transit RMS, and then these fixed apertures were applied to all subsequent images. Frame-to-frame shifts were corrected using cross-correlation tracking (`TRACK = 'x'`) within DEFOT. This allowed the centroids of the target and comparison stars to be accurately followed throughout the observations. During most runs, the Danish telescope's tracking system maintained an RMS pointing stability of approximately 1 to 2 pixels over the course of a typical observing sequence lasting 2 to 4 hours.

The observations were deliberately defocused (Southworth et al. 2009b,a; Tregloan-Reed et al. 2013) to enhance photometric precision by spreading starlight across more pixels and minimizing flat-fielding noise; the additional optimization of aperture size for each dataset helped account for variations in PSF shape and atmospheric seeing, resulting in improved signal-to-noise ratios and more precise light curves. We also verified that the variation in aperture size between datasets did not systematically affect residual scatter.

Differential photometry was constructed by comparing WASP-19's flux against a combined ensemble of reference stars. The number of comparison stars typically ranged from three to six, depending on the field quality. Initially, we used a first-order polynomial ($N = 1$) to account for the dominant variations. However, due to the limited number of suitable comparison stars available, we found it necessary to increase the polynomial order ($N > 1$) to better model and remove additional systematic effects present in the data. The best-fitting comparison star weights were typically close to unity, as expected for observations dominated by Poisson noise in the photon counts.

[1] https://skottfelt.dk/dk154/index.php?title=Main_Page

### 3.2. TESS space telescope

The NASA Transiting Exoplanet Survey Satellite (TESS; Ricker et al. 2015) observed transits of the exoplanet WASP-19 b during its observations in TESS Sector 9 (2019-03-01 to 2019-03-25), Sector 36 (2021-03-07 to 2021-04-01), Sector 62 (2023-02-12 to 2023-03-10), Sector 63 (2023-03-10 to 2023-04-06), Sector 89 (2025-02-11 to 2025-03-12) and Sector 90 (2025-03-12 to 2025-04-09). The TESS observations were conducted with a two-minute cadence, providing high-precision photometric data on the transit events of this hot-Jupiter exoplanet. To analyze this extensive dataset, we utilized the TESS presearch data conditioning (PDC) photometry (Stumpe et al. 2014) that had been processed through the Science Processing Operations Center (SPOC) pipeline (Jenkins et al. 2016). This PDC photometry has been corrected for instrumental systematic effects, enabling a robust analysis of the transit light curves to study the orbital and atmospheric properties of WASP-19 b. A total of 183 transits were observed by TESS, spanning six sectors with an average of 30 transits per sector. These TESS transits were then phase folded to an orbital period found from running a periodogram analysis separately for each sector, resulting in one representative transit light curve per sector. Refer to Fig. A.1 for TESS transits.

### 3.3. Literature data

WASP-19 b has been extensively observed and studied by various research teams over the years. We have collected an additional 43 complete transit light curves from previously published works: Hebb et al. 2010; Albrecht et al. 2012; Anderson et al. 2013; Tregloan-Reed et al. 2013; Hellier et al. 2011; Mancini et al. 2013; Lendl et al. 2013; Dragomir et al. 2011; Bean et al.

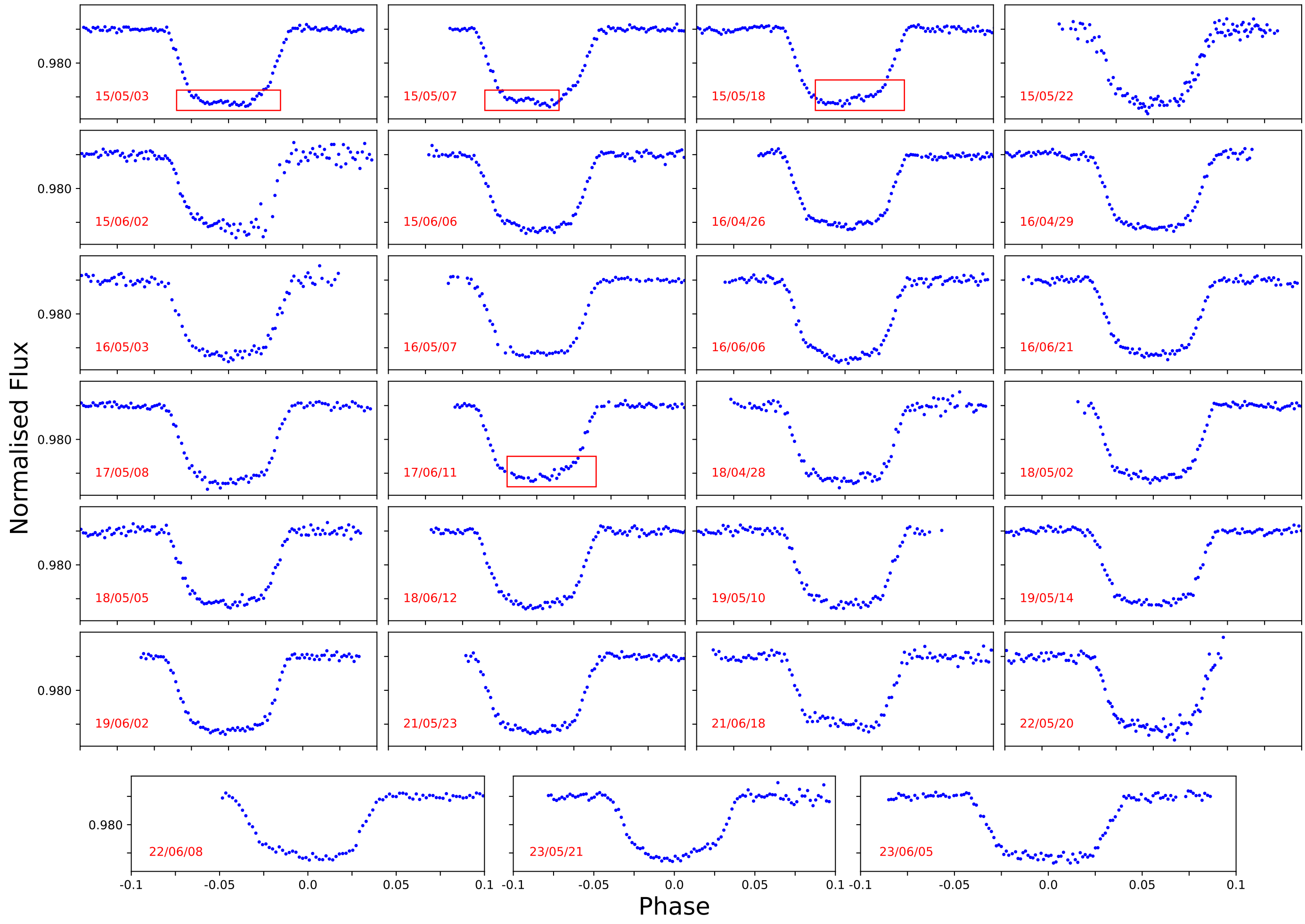

**Fig. 1.** Observations of 27 transits were conducted from 2015 to 2023 utilizing the DFOSC instrument on the Danish Telescope at La Silla Observatory. Transits that exhibit possible starspots are enclosed within rectangular markers. The observation dates follow the YY/MM/DD format.

2013; Espinoza et al. 2019; Sedaghati et al. 2017; Patra et al. 2020; Sedaghati et al. 2015a; Petrucci et al. 2020. These previously published observations will allow us to expand the temporal baseline of our investigation and provide crucial insights into the long-term evolution of the WASP-19 system. For mid-transit timing information, see Table A.2.

## 4. Light curve analysis

### *4.1. Starspot modeling and analysis*

To model the influence of stellar activity on transit light curves, we used the `PRISM` (Planetary Retrospective Integrated Starspot Model) code[2] (Tregloan-Reed et al. 2013). This tool enables the simultaneous modeling of planetary transits and potential starspot crossings by pixelating the stellar disk and simulating flux variations caused by occulted spots. `PRISM` is coupled with the `GEMC` (Genetic Evolution Markov Chain) optimization algorithm, which merges genetic algorithms with Markov chain Monte Carlo (MCMC) techniques. This hybrid framework performs global optimization in the first stage and then refines the best-fit parameters in a Bayesian context during the second stage via differential evolution Markov chain Monte Carlo (`DEMC`).

[2] `https://github.com/JTregloanReed/PRISM_GEMC`

The modeling assumes a quadratic limb-darkening law and can accommodate circular or eccentric orbits. The photometric transit model is defined by the following key parameters: the sum of fractional radii ($r_* + r_p$), the ratio of radii ($k = R_p/R_* = r_p/r_*$), the orbital inclination ($i$), the limb-darkening coefficients ($u_1$, $u_2$), and the mid-transit time ($T_0$). Starspots are modeled with four additional parameters: the spot latitude and longitude ($\theta$, $\phi$), which are directly fitted, the angular radius ($r_{\rm spot}$), and the spot contrast ($\rho$), defined as the ratio of the surface brightness of the starspot to that of the surrounding photosphere.

We applied this modeling framework to the Danish telescope data, where several light curves exhibited anomalies consistent with spot-crossing events (see Fig. 2). For those events, we fit the light curves using `PRISM+GEMC` to characterize the spot properties. Similarly, where applicable, we analyzed TESS light curves and literature data (e.g., WASP-19: Tregloan-Reed et al. (2013); WASP-6: Tregloan-Reed et al. (2015); HAT-P-32: Tregloan-Reed et al. (2018); TESS light curves: Tregloan-Reed & Unda-Sanzana 2019, 2021) using the same approach.

### *4.2. Reanalysis of literature light curve data for potential starspots*

To ensure homogeneity in the analysis, previously published light curves from Lendl et al. (2013), Patra et al. (2020), and

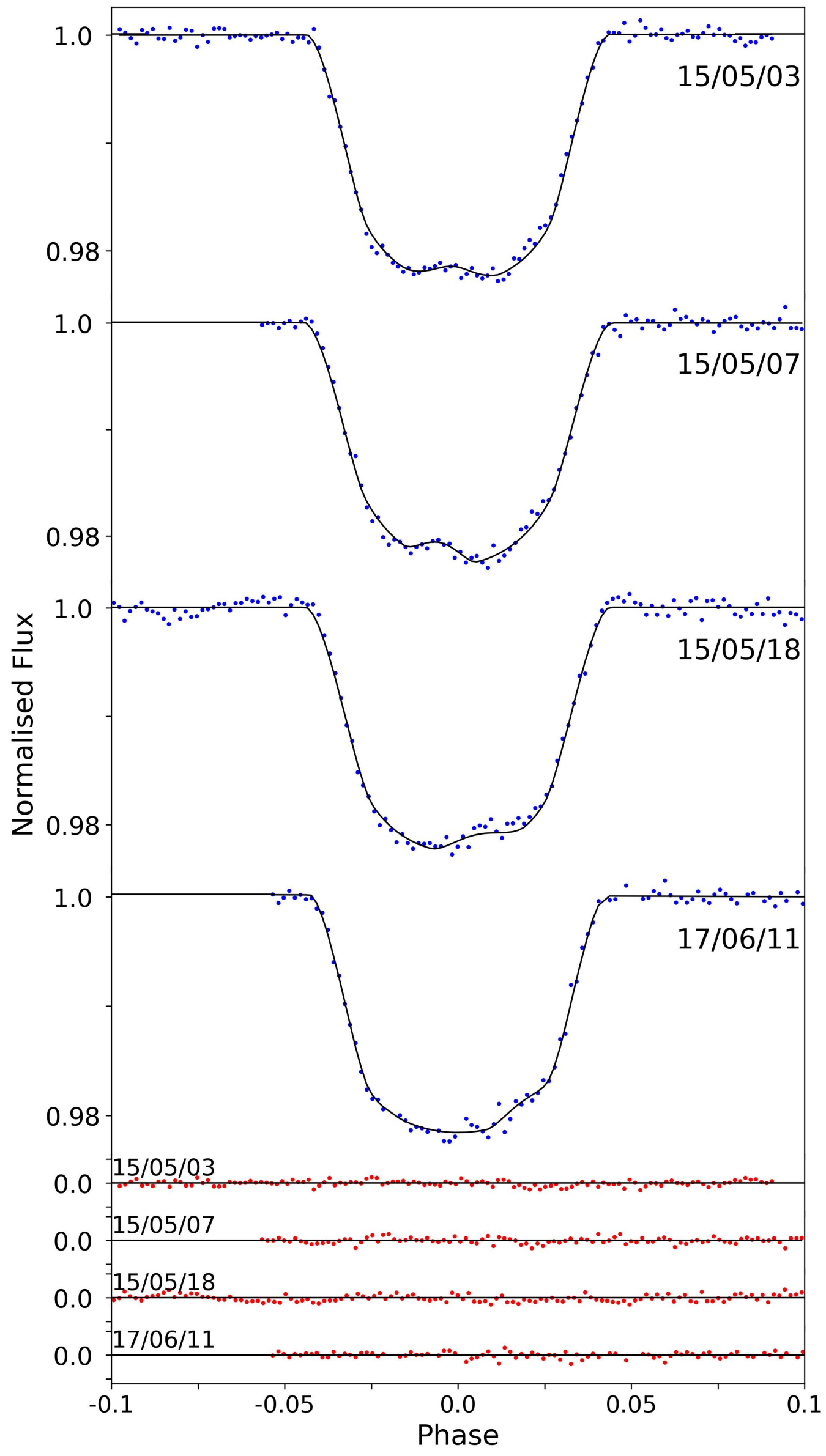

**Fig. 2.** Light curves of transits observed with the Danish telescope, with starspots modeled using PRISM+GEMC, along with the corresponding residuals.

Petrucci et al. (2020) were reanalyzed using the PRISM+GEMC modeling framework. By reprocessing these observations with a uniform modeling approach, potential systematics arising from different reduction and modeling techniques in the literature are minimized, thereby enhancing the reliability and comparability of the derived parameters.

A comprehensive summary of the identified and modeled starspots, including their parameters and uncertainties, is presented in Table C.2. This table provides details on the starspot locations, radii, contrast ratios, and associated uncertainties, facilitating direct comparisons between different observational datasets and improving the robustness of our conclusions regarding the stellar variability in the WASP-19 system.

### *4.3. Geometric and physical parameters results*

Each light curve from the Danish telescope was modeled independently, along with transit light curves compiled from the literature (see Sect. 4.2). The TESS observations were treated separately, with the transits of WASP-19 extracted and fitted individually due to their high cadence and large number.

After individual modeling, the photometric parameters from all datasets were combined using a weighted mean approach to obtain final representative values for the system. These combined results provide robust constraints on the physical configuration of the WASP-19 system. The final photometric parameters, along with comparative values from previous studies, are presented in Table 1.

**Table 1.** Combined photometric parameters for WASP-19 compared with previous studies.

| | $r_p/r_*$ | $r_* + r_p$ | $i$ (deg) |
|---|---|---|---|
| This work | **0.1415 ± 0.0014** | **0.3210 ± 0.0015** | **79.61 ± 0.0984** |
| H10 | 0.1424 ± 0.02 | 0.3222 ± 0.03[a] | 80.7 ± 0.7 |
| AL12 | 0.1425 ± 0.0014 | 0.3432 ± 0.002[a] | 78.02 ± 2 |
| AN13 | 0.1431 ± 0.0015 | 0.3356 ± 0.0032[a] | 79.42 ± 0.39 |
| TR13 | 0.1428 ± 0.0006 | 0.3301 ± 0.0019[a] | 78.94 ± 0.23 |
| H11 | $0.1433^{+0.0018}_{-0.0030}$ | $0.3282^{+0.0070}_{-0.0063}$ | $78.85^{+0.55}_{-0.66}$ |
| S17 | $0.1416^{+0.0019}_{-0.0018}$ | $0.3205^{+0.0038}_{-0.0030}$ | $80.36^{+0.76}_{-0.81}$ |
| P20 | 0.145 ± 0.006 | 0.3245 ± 0.00147[a] | 79.3 ± 1.3 |

**Note.** The values found by **H10** (Hebb et al. 2010); **AL12** (Albrecht et al. 2012); **AN13** (Anderson et al. 2013); **TR13** (Tregloan-Reed et al. 2013); **H11** (Hellier et al. 2011); **S17** (Sedaghati et al. 2017) and **P20** (Petrucci et al. 2020). The photometric parameters are the weighted means from the datasets, which have measured uncertainties. The sum of the fractional radii from the literature was calculated using the respective values for $R_p/R_*$ and $a/R_*$.

To derive the physical parameters of the WASP-19 system, we employed a model-independent method that does not rely on stellar evolutionary models such as MESA (Modules for Experiments in Stellar Astrophysics) or isochrone fitting. Instead, we used a direct estimate of the stellar radius based on the Infrared Flux Method IRFM (Blackwell & Shallis 1977), combined with Gaia DR3 parallaxes and broadband spectral energy distributions. This approach follows the methodology of Goswamy et al. (2024), who derived homogeneous radii for all WASP host stars.

This approach is particularly advantageous because it minimizes the systematic uncertainties and assumptions inherent in model-based techniques. For example, isochrone fitting typically requires prior knowledge of the star's age, metallicity, and evolutionary state, factors that can bias the derived planetary parameters. In contrast, a model-independent method allows us to use purely geometric and dynamical constraints from the light curve and radial velocity data, leading to more robust and transparent results.

Adopting a stellar radius of $1.003^{+0.007}_{-0.007}$ $R_\odot$ from Goswamy et al. (2024), and following the procedure outlined in Morrell et al. (2026), we derived the complete set of stellar and planetary parameters and presented them in Table 2.

## 5. Timing analysis

To measure the TTVs, we included all available transit light curves, both from this work and from previously published datasets, and reanalyzed them homogeneously using the PRISM+GEMC modeling framework. In total, 114 transit events were modeled to extract self-consistent mid-transit times, excluding any secondary eclipse (occultation) timings reported in the literature. The derived mid-transit times and their residuals for each of these events are listed in Table A.2.

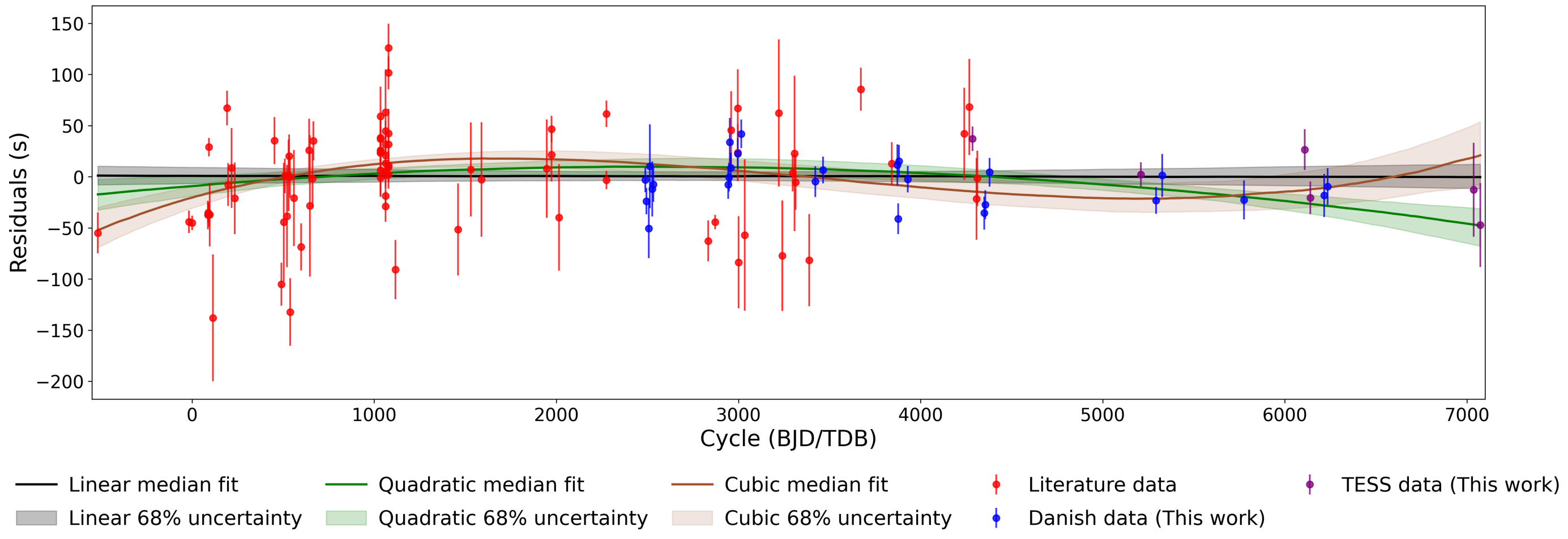

**Fig. 3.** Transit-timing residuals of WASP-19 b with respect to the best-fitting linear ephemeris. The red, blue, and purple points represent literature, Danish, and TESS data, respectively, with their associated timing uncertainties. The solid black, green, and brown curves correspond to the median fits from the linear, quadratic, and cubic polynomial models, respectively. Shaded regions indicate the 68 % confidence intervals derived from 20000 bootstrap realizations of each model's MCMC posterior distribution.

**Table 2.** Physical parameters of the WASP-19 system.

| Parameter | Value |
|---|---|
| Stellar mass ($M_{\odot}$) | $0.914^{+0.019}_{-0.019}$ |
| Stellar radius ($R_{\odot}$) | $1.003^{+0.007}_{-0.007}$ |
| Stellar surface gravity (cgs) | $4.396^{+0.004}_{-0.004}$ |
| Stellar density ($\rho_{\odot}$) | $0.906^{+0.004}_{-0.004}$ |
| Effective temperature (K) | $5460^{+88}_{-89}$ |
| Planet mass ($M_{Jup}$) | $1.171^{+0.125}_{-0.123}$ |
| Planet radius ($R_{Jup}$) | $1.388^{+0.009}_{-0.009}$ |
| Planet surface gravity ($\rho_{Jup}$) | $0.437^{+0.046}_{-0.046}$ |
| Equilibrium temperature (K) | 2070.39 ± 33 |
| Semimajor axis (AU) | $0.016^{+0.0001}_{-0.0001}$ |
| Inclination (degrees) | 79.61 ± 0.10 |
| Eccentricity ($e$) | 0.0 |

### 5.1. *Fitting the mid-transit timing data*

To investigate the presence of orbital decay in the WASP-19 system, we fitted linear, quadratic, and cubic polynomial models to the residuals between the observed mid-transit times and those predicted by a constant-period ephemeris. Specifically, we first used best-fit linear ephemeris to compute expected transit times under the assumption of a strictly periodic, unperturbed orbit. The residuals, i.e., the differences between the observed and predicted transit times, were then modeled as a function of epoch number $E$ using increasingly complex polynomial representations.

The linear model represents a constant orbital period and a fixed reference mid-transit time, which includes two free parameters ($k_F = 2$)

$$T_{\mathrm{tra}}(E) = T_0 + E \times P \,, \qquad (1)$$

where $T_0$ is the reference mid-transit time and $P$ is the orbital period. The quadratic model represents a scenario in which the orbital period changes uniformly with time, introducing a constant period derivative. This form of evolution is the expected signature of tidal dissipation; however, other processes such as long-term apsidal precession, dynamical perturbations from an unseen companion, or secular variations induced by stellar magnetic activity cycles or the Applegate mechanism can produce comparable curvature in the timing residuals. This involves three free parameters ($k_F = 3$)

$$T_{\mathrm{tra}}(E) = T_0 + E \times P + \frac{1}{2} \times \frac{dP}{dE} \times E^2 \,. \qquad (2)$$

Finally, a cubic model was included, which extends the quadratic approach by allowing for an additional degree of freedom, representing variations in the rate of period change itself. It involves four free parameters ($k_F = 4$), making it more flexible in capturing complex changes in the orbital period, such as higher-order effects due to tidal dissipation or interactions with other bodies (e.g., additional planets or stellar companions). A positive cubic term could suggest that the rate of orbital decay is slowing down or, conversely, accelerating, depending on its sign and magnitude.

Two widely employed statistical tools to compare models of diverse levels of complexity are the Akaike information criterion (AIC) and the Bayesian information criterion (BIC). These criteria balance the goodness of fit, quantified by the chi-squared statistic ($\chi^2$), against the complexity of the model, represented by the number of free parameters ($k_F$).

The formulas for the AIC and BIC are defined as follows

$$BIC = \chi^2 + k_F \log N_P \,, \qquad (3)$$

$$AIC = \chi^2 + 2k_F \,, \qquad (4)$$

where, $\chi^2$ is the chi-squared statistic representing the goodness of fit, $k_F$ is the number of free parameters in the model, $N_P$ is the number of data points.

The timing dataset is heterogeneous, combining transit times derived from multiple instruments and analyzes over a ∼15-year baseline. As a result, the formal timing uncertainties do not fully capture the observed scatter in the residuals, and the reduced $\chi^2$ values exceed unity for all the tested ephemeris models. Table 3, illustrated in Fig. 3, shows that the reduced $\chi^2$ decreases as the ephemeris becomes more complex; however, this behavior alone

**Table 3.** Comparison of linear, quadratic, and cubic ephemeris model for WASP-19 b.

| Quantity | Linear ephemeris | Quadratic ephemeris | Cubic ephemeris |
|---|---|---|---|
| Reference time ($T_0$; BJD(TDB)-2400000) | 55183.167643±0.000061 | 55183.167527±0.000074 | 55183.167368±0.000080 |
| Period ($P_b$; days) | $0.788839059^{+0.000000023}_{-0.000000024}$ | $0.788839224^{+0.000000068}_{-0.000000069}$ | $0.788839703^{+0.000000136}_{-0.000000135}$ |
| Quadratic term ($p1$) | | $-3.1536 \times 10^{-11}\,^{+1.2297\times10^{-11}}_{-1.2321\times10^{-11}}$ | $-2.5574 \times 10^{-10}\,^{+5.7502\times10^{-11}}_{-5.6958\times10^{-11}}$ |
| Cubic term ($p2$) | | | $2.946 \times 10^{-14}\,^{+6.419\times10^{-15}}_{-6.364\times10^{-15}}$ |
| $\chi^2$ | 547.02 | 516.29 | 468.47 |
| $\chi^2_\nu$ | 4.84 | 4.61 | 4.22 |
| AIC | 551.02 | 522.29 | 476.47 |
| BIC | 556.51 | 530.53 | 487.45 |
| r.m.s. of residuals (s) | 48.54 | 47.51 | 46.31 |

**Note.** The models are fitted to the full set of mid-transit times spanning 2008–2025 (Table A.2). The table lists the raw $\chi^2$, reduced $\chi^2_\nu$, Akaike information criterion (AIC), and Bayesian information criterion (BIC) values computed directly from the measured timing uncertainties. While none of the models fully reproduces the observed scatter in the timing residuals, the cubic ephemeris yields the lowest information-criterion values among the tested models.

may simply reflect the increased flexibility of higher-order models rather than an unambiguous physical significance. To assess whether the inclusion of additional parameters is statistically justified, we therefore rely on AIC and BIC as relative model comparison metrics. The BIC evaluates which model is most strongly supported when penalizing complexity, while the AIC identifies the model that optimally balances goodness of fit and parsimony. Both criteria consistently favor cubic ephemeris over linear and quadratic alternatives when applied to the full dataset listed in Table A.2.

Physically, the cubic term captures a slow, non-periodic trend in the orbital evolution of WASP-19 b. Such behavior may reflect a low-order secular process, such as tidal dissipation or apsidal precession, that would not translate into a strictly periodic signal. The resulting curvature in the transit timing residuals thus provides compelling evidence of non-linear orbital evolution within the system (e.g., Maciejewski et al. 2018; Penev et al. 2018; Gomes et al. 2021; Yalçınkaya et al. 2024). This interpretation is consistent with theoretical expectations that tidal dissipation and apsidal motion in short-period hot-Jupiters induce secular, non-periodic drifts in transit timings rather than sinusoidal variations, producing the type of curvature captured by a cubic ephemeris. As discussed previously, other mechanisms can, in principle, also generate higher-order trends. In contrast to previous works (see, e.g., Petrucci et al. 2020) that favored a purely linear ephemeris and found no sign of decay, our bootstrap-based MCMC analysis (see below) supports the cubic model, suggesting that the planet's orbit may be undergoing subtle, measurable changes consistent with long-term dynamical effects.

Model selection robustness against outliers To evaluate whether the detected departure from constant-period timing variations is influenced by the heterogeneous quality of the dataset, we compared model performance for the complete sample (including TESS) and for a subset excluding TESS (see Appendix A). As expected from its higher photometric scatter, the inclusion of TESS increases the values of the overall $\chi^2$ and the information-criterion values. Nevertheless, in both cases, the cubic ephemeris remains consistently favored over the linear and quadratic models, indicating that the long-term curvature in the timing residuals is not driven by the lower-precision TESS points.

To further assess robustness against potentially influential measurements, we performed a series of complementary filtering tests. Removing the most uncertain transit times of 5 % produced only a minor reduction in scatter, while excluding points beyond $1\sigma$ or $2\sigma$ from the best-fitting trend yielded greater improvements in $\chi^2$, AIC, and BIC. In all instances, the cubic model continued to provide the lowest information-criterion scores, demonstrating that no single point or subset governs the inferred curvature.

Finally, we applied a random hold-out cross-validation procedure ($X_{\mathrm{val}}$), in which 20 % of the timings were removed randomly and the remaining 80 % refitted over $10^4$ iterations. Across all realizations, the cubic ephemeris delivered prediction errors with a median absolute deviation comparable to the simpler models while showing the smallest systematic bias. Although the reduced values $\chi^2$ remain greater than unity in these tests, reflecting the intrinsic heterogeneity of the timing dataset and the presence of excess scatter beyond the formal uncertainties quoted, this behavior is consistent between all tested models. The persistence of the cubic preference under cross-validation therefore indicates that the inferred non-linear trend is not an artifact of individual outliers or underestimated uncertainties, but a coherent long-term feature of the data. This confirms that the non-linear trend persists regardless of which measurements are withheld.

### 5.2. Tidal quality factor $Q'_*$

Although tidal decay would, observationally, mean as a quadratic trend in transit timings, our model comparison shows that the quadratic ephemeris is not statistically preferred (see Sect. 5.1). To independently assess whether tidal decay could still be responsible for the observed deviations, we proceed to constrain the modified stellar tidal quality factor. The tidal quality factor ($Q_*$) is a measure of the efficiency of the dissipation of tidal energy (Ogilvie 2014). To constrain $Q_*$, we follow the approach outlined by Birkby et al. 2014, which describes the modified tidal quality factor as

$$Q'_* = \frac{3}{2}\frac{Q_*}{k_2}, \tag{5}$$

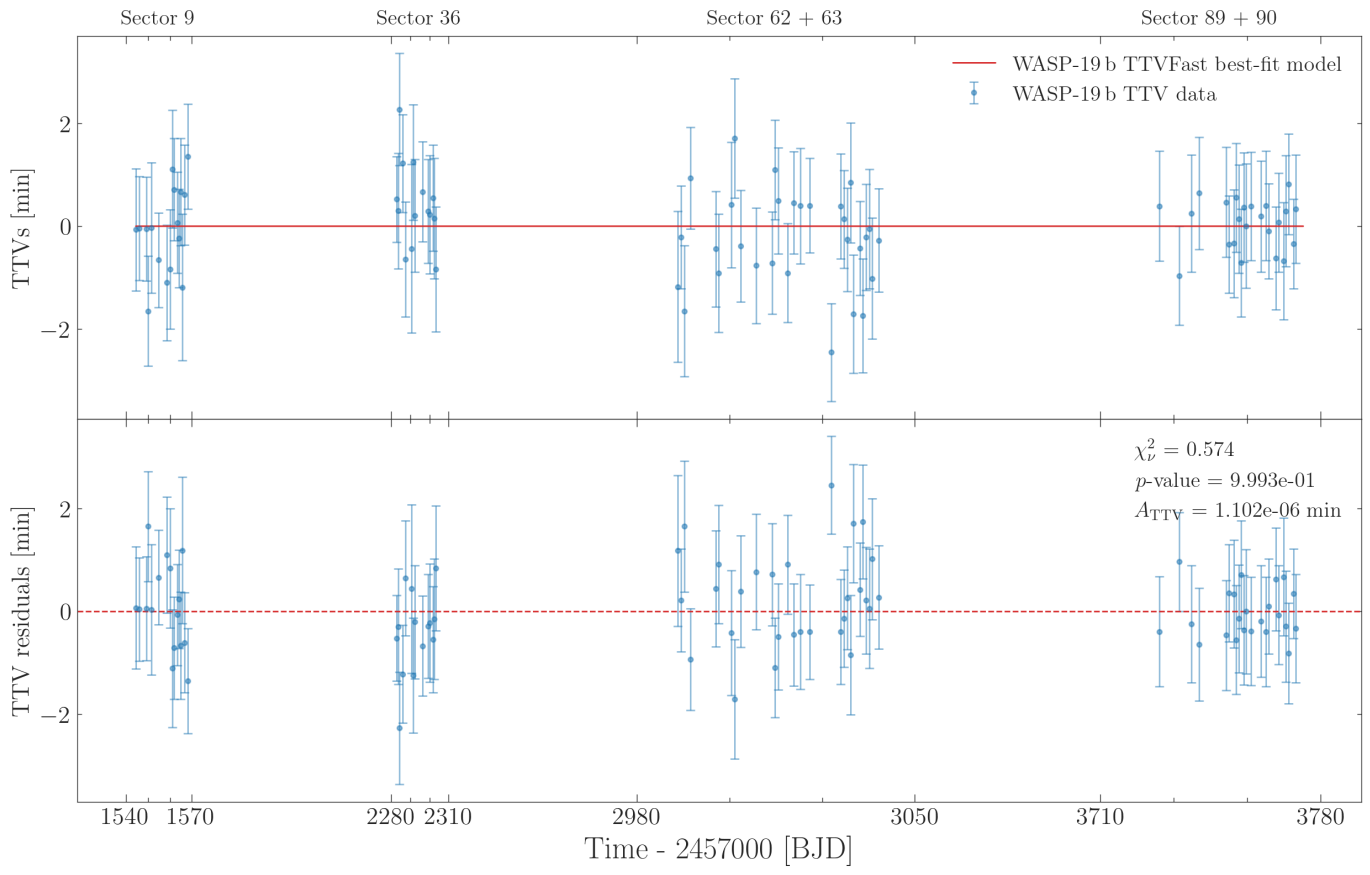

**Fig. 4.** Measured TTVs derived from `allesfitter` when modeling TESS observations for WASP-19 b in the upper plot with their residuals illustrated in the lower plot. The `TTVFast` model yields a negligible TTV amplitude of $A_{\rm TTV} = 1.102 \times 10^{-6}$ min.

where $k_2$ is the Love number (Love 1912). The modified tidal quality factor is related to measurable properties of the system through the equation

$$Q'_* = \frac{-27}{8}\left(\frac{M_p}{M_A}\right)\left(\frac{R_A}{a}\right)^5\left(\frac{2\pi}{P_{orb}}\right)\frac{1}{p1}, \tag{6}$$

where $\left(\frac{R_A}{a}\right)$, denoted as $r_A$, represents the fractional radius of the star, and $p1$ is the quadratic coefficient in the ephemeris.

To estimate the tidal quality factor $Q'_*$, one of the critical steps involves determining which model best fits the data: a model in which the orbital period remains constant or a model in which the orbital period changes (indicating potential orbital decay or perturbations). As said in Sect. 5.1, our data support the cubic ephemeris.

To quantify the extent of orbital decay indicated by the quadratic trend, we utilize the best-fit value of the quadratic term $p1$ = -3.1536 × $10^{-11}$ from Table 3. Although the cubic model yields a better statistical fit (AIC and BIC), the quadratic term directly captures the long-term, monotonic decrease in orbital period associated with tidal decay. However, because our full TTV analysis does not detect decay at this level, the decay rate inferred from the quadratic fit must be interpreted as an upper limit. Following the formulation from Birkby et al. (2014), the derived quadratic term provides a lower limit on the stellar modified tidal quality factor of $Q'_* \approx 10^{6.29\pm0.07}$. The uncertainty given reflects propagated errors in $M_A$, $M_p$, and $r_A$. This lower limit is consistent with previous constraints for the hot-Jupiters (e.g., Bonomo et al. 2017).

### 5.3. Further analysis and interpretation of TTV restricted to TESS data

While in the previous section we searched for TTVs across the entire 15-year dataset, here we refine our restricted analysis (Appendix B) by focusing on TESS-only light curves with 82 valid mid-transit times without the impact of starspots, which offer exceptional photometric precision, cadence uniformity, continuous temporal coverage, and less contamination from heterogeneous ground-based timing data.

For this purpose, we first derived individual TTVs from the TESS sectors using the `allesfitter` framework (Günther & Daylan 2019, 2021), which simultaneously models the transit light curve through a GP baseline model using MCMC to obtain statistically robust uncertainties. The measured TTVs were then compared with the synthetic TTV series computed with the $N$-body `TTVFast` (Deck et al. 2014) simulation. In this context, `allesfitter` provides the measured TTVs that consider all possible astrophysical effects (including gravitational perturbation, tidal dissipation, and potential apsidal motion), whereas `TTVFast` isolates the component of TTVs that arises purely from gravitational perturbations. The comparison between the measured TTVs and synthetic TTV series allows us to assess whether additional secular or dissipative effects are required to explain the observed transit timing residuals in Fig. 3.

The resulting `TTVFast` model for the TESS-only dataset is illustrated in Fig. 4. We chose to perform the TTV analysis exclusively on the TESS dataset because the `TTVFast` simulation imposes a hard limit of 5000 mid-transit times per run. Given the short orbital period of WASP-19 (≈0.79 days), this restricts the effective simulation window to ~ 3950 days, which aligns with the TESS baseline but falls short of spanning the full 15-year ground-based dataset. Thus, using TESS alone enables a consistent and uninterrupted TTV analysis within this constraint. The agreement between the measured TTVs and synthetic TTV series is excellent, yielding a reduced chi-square of $\chi^2_\nu$ = 0.574 and a $p$-value of 0.9993. The inferred TTV amplitude, $A_{\rm TTV} = 1.102 \times 10^{-6}$ min, suggests that the TTVs measured by `allesfitter` are consistent with the `TTVFast` model (Hughes & Hase 2010). No systematic trend or coherent deviation is detected across six TESS sectors, implying that WASP-19 b is dynamically isolated and exhibits no measurable TTVs due to gravitational perturbations from additional companions.

To further explore the dynamical stability of an injected perturber in lower-order near mean-resonance with WASP-19 b, we plotted the mean exponential growth factor of nearby orbits (MEGNO; Cincotta & Simó 2000; Goździewski et al. 2001; Borkovits et al. 2003; Hinse et al. 2010) map through a grid search in the parameter space $(P'/P_b,\, M') \in \left([0.1, 9.9],\, [10^{-6}\, M_\oplus, 10^6\, M_\oplus]\right)$ in Fig. 5 to analyze the orbit stability of an injected perturber at sites close to WASP-19 b. The upper mass limit of the perturber is estimated by applying the `TTV2Fast2Furious` code [3] (TTV2F2F; Hadden et al. 2019) to the `TTVFast` simulation. Under the assumption of a coplanar perturber in a circular orbit as the initial conditions, we start the numerical simulation in this putative three-body system by integrating each grid point from the initial conditions for 60,000 yr to fully saturate chaotic regions in the MEGNO map (Hinse et al. 2010). The MEGNO indicator $\langle Y\rangle$ is then computed and illustrated with color bars, where $\langle Y\rangle \approx 2$ indicates quasi-periodic (regular) dynamics and $\langle Y\rangle \to 5$ indicates chaotic (unstable) behavior. For candidates of the injected perturber in the $\langle Y\rangle \approx 2$ region, they are more likely to be recovered.

Our MEGNO map in Fig. 5 rules out the unstable region between the period ratio 2:3 and 3:2, which is consistent with the MEGNO map presented by Cortés-Zuleta et al. (2020). However, our map did not discover the evidence of 5:2 and 3:1 mean-motion resonances. This difference is probably because we derive the perturber's upper mass limit from WASP-19 b's synthetic TTV series caused by gravitational perturbations, while Cortés-Zuleta et al. (2020) directly calculated the RMS statistics of original transit timing residuals containing all astrophysical effects in the O–C plot.

After combining the MEGNO indicators on our map in Fig. 5 with the RV constraints in the region where $P' < 20$ d (Bern-abò et al. 2024), it turns out that only near 1:2 and near 2:1 resonances with an upper mass limit $\sim 10\,M_\oplus$ are possible for a stable near-resonant perturber. However, since Bernabò et al. (2024) have already excluded the $P' < 20$ d region through a $\chi^2$ analysis by injecting a perturber to their simulation, we do not prefer the existence of a lower-order near-resonant perturber.

[3] https://github.com/shadden/TTV2Fast2Furious

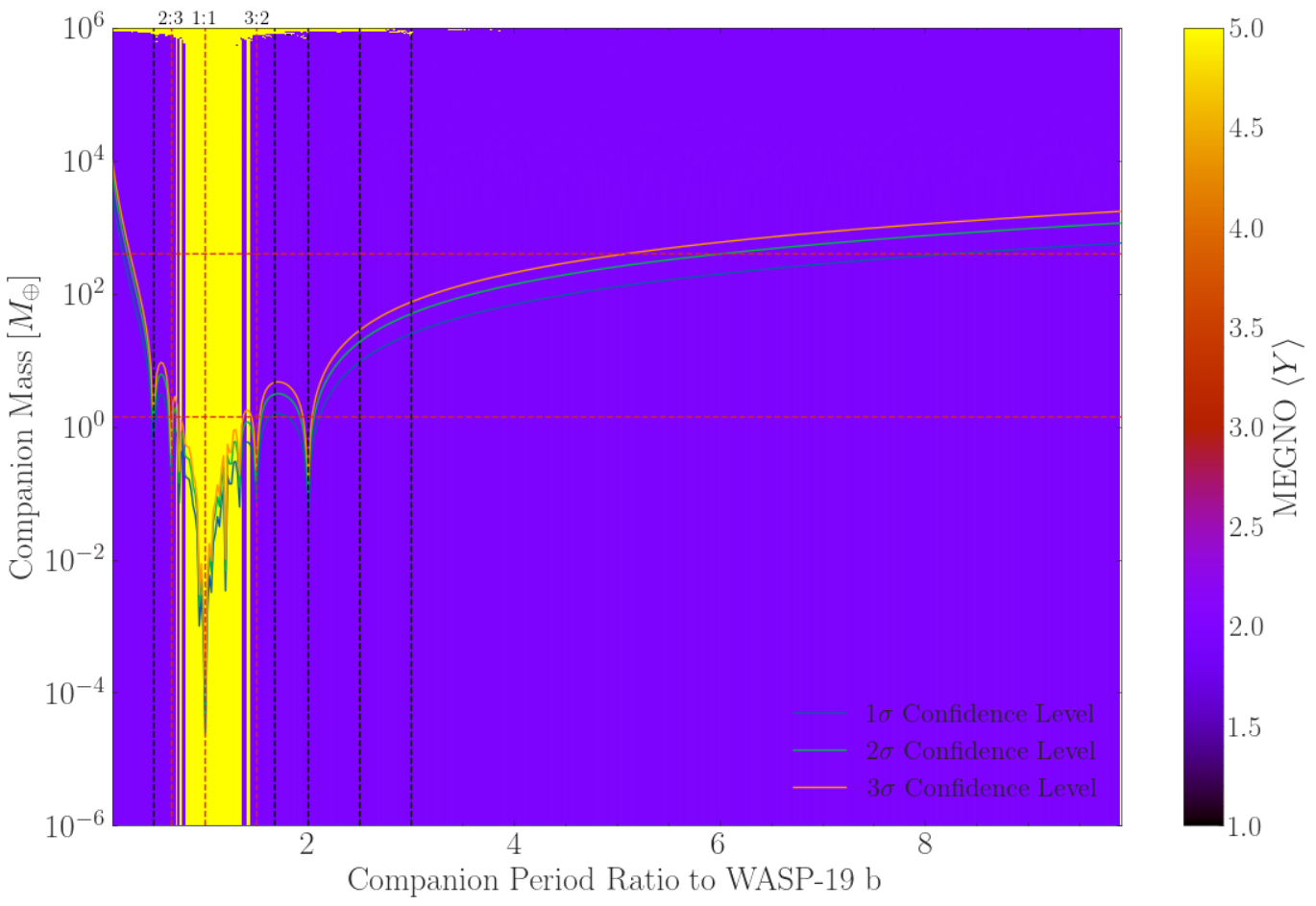

**Fig. 5.** MEGNO stability map depicting the upper mass limit (within $\pm 1\sigma$ (68.3%), $\pm 2\sigma$ (95.4%), and $\pm 3\sigma$ (99.7%) confidence intervals) of an injected hypothetical perturber in the WASP-19 system. The red horizontal dashed lines represent the perturber's mass range between $M = 400\,M_\oplus$ and $M = 1.4\,M_\oplus$ determined from the RV constraints (Bernabò et al. 2024). The red vertical dashed lines from left to right represent the perturber's unstable orbits in 2:3, 1:1, and 3:2 resonances with WASP-19 b determined in this paper, while the black vertical dashed lines from left to right represent the perturber's stable orbits in 1:2, 5:3, 2:1, 5:2, and 3:1 resonances with WASP-19 b found by Cortés-Zuleta et al. (2020).

For a higher-order near resonant exterior perturber, when we are using the `TTV2F2F` code to derive its mass-eccentricity constraints, the contour plot is not very constraining, revealing that there is no strong evidence of the existence of an unseen perturber in the WASP-19 system. Therefore, we conclude that the more favorable cubic model with a smaller $\chi^2_\nu$ cannot be attributed to an unseen perturber.

### 5.4. Constraints on apsidal precession and internal structure of WASP-19 b

To further test the hypothesis that long-term deviations from a linear ephemeris are driven by apsidal precession, we applied the classical precession model described in Ragozzine & Wolf (2009). We modeled the observed mid-transit times using the following expression for an apsidally precessing system

$$T(N) = T_0 + NP_s - \frac{eP_s}{\pi}\cos\left(\omega_0 + N\dot{\omega}\right), \quad (7)$$

where $T_0$ is the reference transit time, $P_s$ is the sidereal period, $e$ is the orbital eccentricity, $\omega_0$ is the argument of the periastron at epoch zero, and $\dot{\omega}$ is the apsidal precession rate in radians per orbit. We fixed the eccentricity at $e = 0.00172$, following the analysis of Bernabò et al. (2024), who derived this value from the combined RV and photometric constraints.

Using non-linear least squares optimization, we fit this model to our compiled transit mid-times, which span over 15 years. The resulting best-fit apsidal precession rate was $\dot{\omega}_{\rm obs} = (1.00 \pm 0.12) \times 10^{-4}$ rad/orbit.

To interpret this observationally derived value, we computed the expected contribution from general relativity using the equation below, and by substituting the system parameters, we found

$$\dot{\omega}_{\rm GR} = \frac{6\pi G M_*}{a(1-e^2)c^2} = 1.05 \times 10^{-5}\ \text{rad/orbit}. \quad (8)$$

Subtracting this from the total observed rate yields the tidal component of the precession

$$\dot{\omega}_{\rm tide} = \dot{\omega}_{\rm obs} - \dot{\omega}_{\rm GR} = 8.95 \times 10^{-5}\ \text{rad/orbit}. \quad (9)$$

The planetary Love number $k_{2p}$, which characterizes the planet's interior response to tidal forces, can then be derived using

$$\dot{\omega}_{\rm tide} = \frac{15}{2} k_{2p} \left(\frac{R_p}{a}\right)^5 \left(\frac{M_*}{M_p}\right) \frac{1}{(1-e^2)^5}. \quad (10)$$

Rearranging, we solved for $k_{2p}$

$$k_{2p} = \frac{\dot{\omega}_{\rm tide}}{\frac{15}{2}\left(\frac{R_p}{a}\right)^5 \left(\frac{M_*}{M_p}\right) \frac{1}{(1-e^2)^5}} = 0.107 \pm 0.08. \quad (11)$$

This value agrees well with the result of Bernabò et al. (2024), who reported $k_{2p} = 0.20^{+0.02}_{-0.03}$ based on a joint analysis of transits, occultations, and radial velocity data (see their Sect. 6.1). Nevertheless, the two estimates are consistent within $2\sigma$ and reinforce the interpretation that WASP-19 b experiences an observable apsidal precession due to tidal distortion.

Combined with the statistical evidence of a cubic ephemeris (see Sect. 5.1), this dynamical modeling reinforces the view that the long-term orbital evolution of WASP-19 b is not driven by orbital decay, but by apsidal motion resulting from both internal planetary structure and possible secular forcing from a wide-orbit stellar companion identified in Bernabò et al. (2024).

## 6. Transmission spectrum analysis

WASP-19 b has been extensively studied by several researchers (e.g., Mancini et al. 2013; Mandell et al. 2013; Sedaghati et al. 2015a). However, only a few have taken stellar anomalies into account, which can significantly impact the accuracy of the results. By modeling and correcting our data homogeneously for the effects of starspots and other stellar activity in the light curve, we can substantially improve the precision of the measured planet-to-star radius ratio, with uncertainties reduced by up to a factor of two compared to earlier studies. For example, we report a refined value of $R_{\rm p}/R_\star = 0.1415 \pm 0.0014$, compared to previously published uncertainties ranging from $\pm 0.002$ to $\pm 0.006$ (e.g., H10, AL12, P20; see our Table 1). Such precision is crucial when interpreting atmospheric absorption features in transmission spectra, especially in hot-Jupiters like WASP-19 b, where stellar activity is known to distort transit depths (Rackham et al. 2018; Oshagh et al. 2014). In our analysis of WASP-19 b, we incorporated starspot modeling into the light curve fitting process. This approach allowed us to refine the transit depth measurements, enhancing our ability to detect subtle atmospheric absorption features. Correcting for stellar activity not only reduces systematic errors, but also increases sensitivity to important atmospheric signals, such as molecular absorption bands, which may otherwise be obscured by noise.

### 6.1. Photometric spectrum

In our study of WASP-19 b, a strongly irradiated hot-Jupiter, we used photometric data to reconstruct a low-resolution photometric transmission spectrum. Since our objective was closely tied to the orbital decay study, we did not incorporate occultation data, which would otherwise be necessary to characterize the planet's dayside emission. Although such data can complement atmospheric characterization, especially for retrieving temperature profiles and detecting thermal inversions, their inclusion is not critical for probing transmission features, which are dominated by the planetary limb. Instead, we focused exclusively on transit observations to enhance the reliability of our constraints on orbital precession and the transmission spectrum derived purely from primary transits. We employed light curves obtained from the Danish telescope in the R and I filters, TESS in the R filter, along with multi-wavelength light curves from the literature across different passbands (g, J, H, K). These light curves allowed us to investigate the atmospheric characteristics on the nightside of WASP-19 b. To ensure the accuracy of our analysis, we excluded incomplete or unreliable light curves, as discussed in Sect. 3.1, and focused solely on complete datasets. Using the PRISM+GEMC algorithm, we analyzed each light curve to compute the planet-to-star radius ratio ($R_p/R_*$) for each passband. During this process, we kept other photometric parameters constant, using the best-fitting values determined from the overall light-curve fitting. This method ensured that any observed variations in the radius ratio were genuine and not influenced by systematic errors or instrumental effects.

For each of the six passbands, we calculated the $R_p/R_*$ values and then computed a weighted mean for each passband to increase the precision of our estimates. These weighted averages formed the basis for our reconstructed photometric transmission spectrum. Although this multi-wavelength photometry is lower in resolution than traditional transmission spectroscopy, it provides valuable insight into the composition of WASP-19 b's nightside atmosphere.

### 6.2. Variation of planetary radius with the wavelength

To ensure a consistent transmission spectrum across heterogeneous datasets, we applied small additive transit-depth offsets to each instrument, derived by aligning all measurements with the high-S/N *Hubble* Space Telescope (HST) STIS+WFC3 baseline (Huitson et al. 2013) through weighted interpolation and bootstrap-estimated uncertainties with final offset values listed in Table C (see Appendix C for full details of the baseline correction procedure).

Figure 6 shows the variation of $k$ as a function of wavelength, obtained from multiple datasets, including Danish-DFOSC instrument (R and I band), TESS space telescope (R band), GROND instrument covering multiple bands (Sloan g′, Sloan r′, Sloan i′, J, H, K) from Mancini et al. (2013) and MMIRS instrument (J, H, K bands) from Bean et al. (2013). Additionally, spectroscopic data points from Sedaghati et al. (2015a) and Huitson et al. (2013) were included to increase the accuracy of the model. These datasets are compared against two atmospheric retrieval models, one that includes metal oxides: titanium oxide (TiO) and vanadium oxide (VO), and another that excludes these species.

Retrievals were performed using POSEIDON (MacDonald 2023; MacDonald & Madhusudhan 2017), a Bayesian atmospheric retrieval framework with 2000 live points. Best-fit retrieval assuming opacity sources from Na, K, and $H_2O$, excluding TiO/VO (Orange line) predominately explains the observed spectrum of sodium (Na) at ∼0.59 $\mu$m, potassium (K) at ∼0.77 $\mu$m, water vapor ($H_2O$) at ∼1.4 $\mu$m, and Rayleigh scattering at blue optical wavelength in the $He$-$H_2$ dominated atmosphere. Given the extreme irradiation of WASP-19 b, the presence of metal oxides such as titanium oxide (TiO) and vanadium oxide (VO) is also anticipated. The best-fit incorporating TiO and VO opacities in the above spectrum are shown as the purple line. Metal oxides contribute significantly to atmospheric opacities, leading to stronger absorption features in the UV-optical wavelength (∼0.4 - 1.0 $\mu$m).

The updated retrieval analysis based on the offset-corrected transmission spectrum further strengthens the statistical preference for an atmosphere without TiO/VO. The TiO/VO-free model yields a higher Bayesian evidence of $\ln Z = 354.29 \pm 0.19$, compared to $\ln Z = 350.37 \pm 0.22$ for the model including TiO and VO. This difference ($\Delta \ln Z \simeq 4$) represents strong statistical support for the TiO/VO-free scenario. In addition, the reduced chi-squared value improves significantly: the TiO/VO-free model achieves $\chi^2_\nu = 1.85$ (with $\chi^2 = 96.17$ for 52 degrees of freedom), whereas the model including TiO/VO yields a slightly poorer fit with $\chi^2_\nu = 1.89$ ($\chi^2 = 98.40$). These improvements, which are substantially better than our previous pre-offset results ($\chi^2_\nu = 2.74$), demonstrate that homogenizing the heterogeneous datasets through offset correction reduces inter-instrument discontinuities and prevents spurious spectral gradients from being misinterpreted as TiO/VO absorption. Consequently, the offset-corrected retrieval provides a more coherent and physically consistent interpretation of the planetary atmosphere, favoring the presence of Na, K, $H_2O$, and Rayleigh scattering, while offering less compelling statistical evidence of TiO/VO in the transmission spectrum of WASP-19 b.

However, previous high-resolution spectroscopic studies, such as Sedaghati et al. (2017) using VLT-FORS2, reported a tentative detection of TiO features in the transmission spectrum of WASP-19 b. This discrepancy likely arises due to the difference in the spectral resolution and analysis approach. Sedaghati et al. employed differential spectroscopy using FORS2, which provides significantly higher spectral resolution and is capable of isolating narrowband features such as TiO absorption bands in the optical range. In contrast, the photometric transmission spectrum used in our study, though covering a wide wavelength range, relies on broadband photometry with limited spectral resolution.

Such photometric data tend to dilute or smooth out narrow spectral features, making it challenging to resolve molecular opacities like TiO/VO, particularly in the presence of instrumental systematics or stellar heterogeneities. Although we incorporated the spectroscopic data from Sedaghati et al. (2015a) into our combined spectrum, the retrieval model still favors a solution without TiO/VO. This likely reflects the dominant influence of the lower-resolution photometric datasets included in the analysis. In addition, to complement the photometric dataset, synthetic photometric points were generated using the SVO SpecPhot[4] tool by uploading the retrieved model spectra of WASP-19 b (spanning 0.3–2.5,$\mu$m) and integrating them into UV–optical filters: Rosetta/OSIRIS_WAC.F71, F81, UV375, HST/ACS_HRC.F435W, and F475W to obtain flux-weighted transit depths at the corresponding effective wavelengths. These synthetic points (plotted as colored squares in Fig. 6) represent where future or existing photometric observations would probe each model retrieved. They serve to extend the model predic-tions into wavelength regions with limited observational coverage, providing a continuous theoretical reference that highlights the expected spectral behavior of both the TiO/VO-included and TiO/VO-free retrievals.

[4] https://svo2.cab.inta-csic.es/theory/specphot

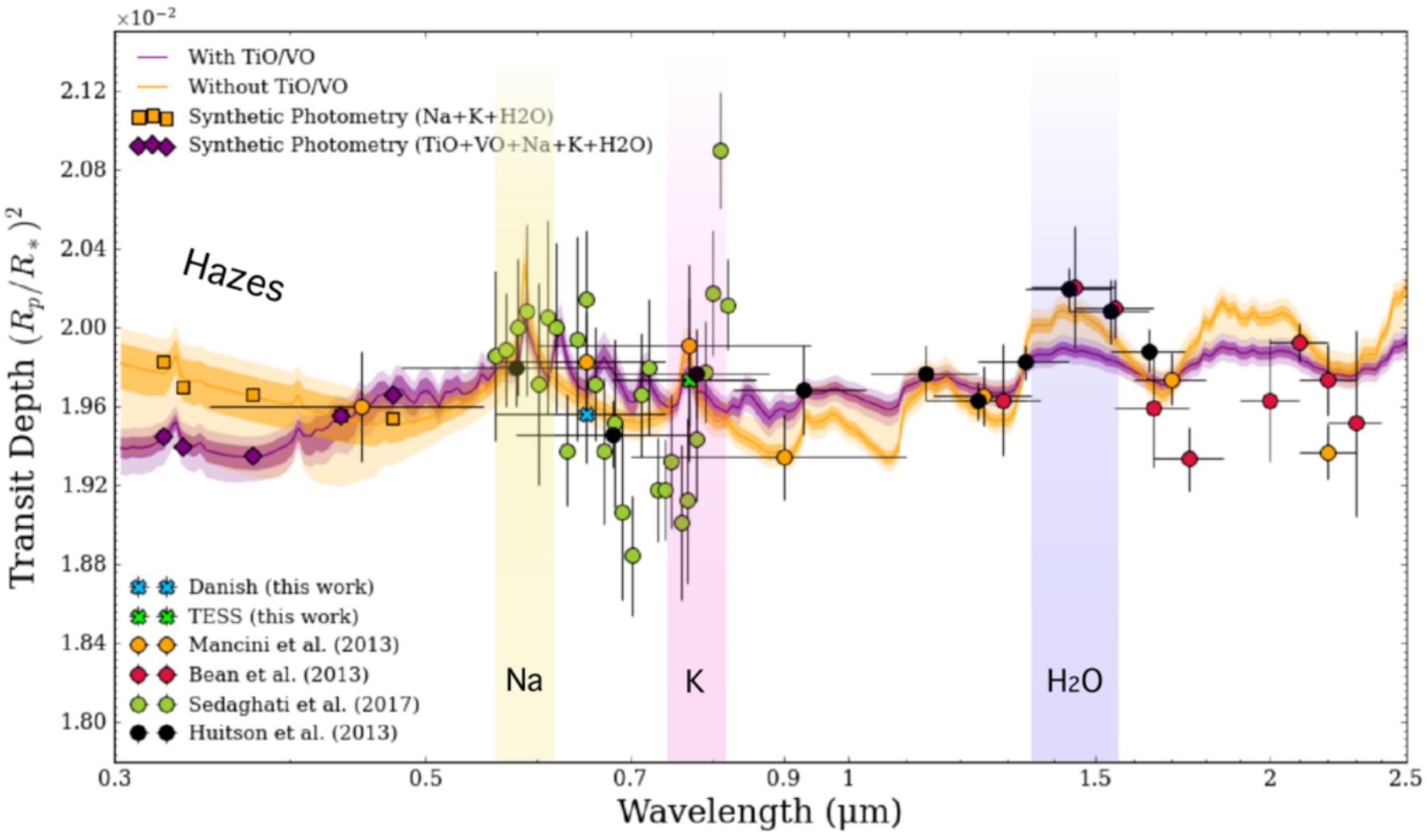

**Fig. 6.** Variation of the planetary radius, expressed as transit depth, as a function of wavelength. Sky-blue points correspond to Danish-DFOSC (this work), lime to TESS (this work), orange to GROND (Mancini et al. 2013), crimson to MMIRS (Bean et al. 2013), green to (Sedaghati et al. 2015a), and black to (Huitson et al. 2013). Vertical error bars represent measurement uncertainties, while horizontal bars indicate the full width half maximum (FWHM) of the respective passbands. Synthetic photometric points were computed using the SVO `SpecPhot` tool to obtain flux-weighted transit depths at the effective wavelengths of the UV–optical filters. Retrievals were performed with `POSEIDON` including the opacities of Na, K, TiO, VO, and $H_2O$. The orange curve shows the median best-fit model excluding TiO and VO, with $1\sigma$ and $2\sigma$ confidence intervals. The purple curve shows the median best-fit model including TiO and VO, with corresponding uncertainties. See Sect. 6.2 for details.

Overall, our result does not refute the possibility of TiO/VO in WASP-19 b's atmosphere, but rather highlights the limitations of broadband photometry in confidently detecting such species. High-resolution, spectrally resolved follow-up observations are essential to confirm the presence and abundance of metal oxides. Although our retrieval analysis does not provide strong evidence of the presence of TiO/VO, it is important to consider that atmospheric composition may not be static. Beyond observational limitations, the dynamical evolution of the planet's orbit itself can significantly affect atmospheric structure and chemistry (Komacek & Showman 2020; May & Rauscher 2021). In hot-Jupiter systems like WASP-19 b, while orbital decay due to tidal dissipation is a widely expected phenomenon, other processes such as apsidal precession may play an equally critical role.

In the case of WASP-19 b, the presence of apsidal precession is suggested by the cubic trend observed in the transit timing residuals (see Sect. 5.1). This kind of orbital evolution is not only of dynamical interest, but it can also lead to periodic changes in the stellar irradiation received by the planet (Langton & Laughlin 2008; Rauscher & Kempton 2014). These fluctuations can result in time-variable atmospheric temperatures, which in turn influence molecular abundances through changes in chemical equilibrium. Consequently, absorbers such as $H_2O$, TiO, and VO may experience redistribution or dissociation over time, leading to variations in the observed transmission spectrum.

## 7. Results

This study provides a comprehensive analysis of the TTVs and atmospheric properties of the ultra-short-period hot-Jupiter, WASP-19 b. Our main results are as follows:

- Using an extensive dataset spanning more than a decade, we modeled the transit timings of WASP-19 b with linear, quadratic, and cubic ephemeris models to assess potential departures from a constant-period orbit. due to the heterogeneous quality of the timing measurements, the reduced values $\chi^2$ exceed unity for all tested ephemerides. Nevertheless, a relative model comparison using the AIC and BIC consistently favors cubic ephemeris over linear and quadratic models (see Table 3). In a purely mathematical sense, the cubic term represents slow, non-periodic curvature in the observed-minus-calculated (O–C) diagram. Physically, such curvature may arise from low-order secular processes, including tidal dissipation or apsidal precession, rather than from periodic perturbations by additional planets. However, given the modest amplitude of the observed deviations (~50–100 s) relative to the general scatter in the timing data, the current dataset does not provide statistically significant evidence of genuine TTVs. The cubic ephemeris should therefore be interpreted as an empirical description of a long-term trend rather than as a definitive proof of ongoing dynamical perturbations. This conclusion is consistent with theoretical expectations that tidal or precessional effects in short-period systems may produce smooth, low-amplitude drifts over multi-year baselines (e.g., Maciejewski et al. 2018; Penev et al. 2018; Gomes et al. 2021; Yalçınkaya et al. 2024), which can only be robustly confirmed through future high-precision monitoring.
- To verify the robustness of the timing signal, we performed a dedicated analysis using only the six available TESS sectors. Individual TTV were derived using `allesfitter` (Günther & Daylan 2019, 2021), and compared with synthetic TTV series from the N-body simulation `TTVFast` (Deck et al. 2014). In this framework, `allesfitter` provides the observed transit timing residuals (including all astrophysical effects), while `TTVFast` isolates the component of TTVs produced purely by gravitational perturbations. The resulting fit (Fig. 4) shows an excellent agreement between the observed TTV and modeled TTV series ($\chi^2_\nu = 0.574$, $p$-value $= 0.9993$), with a negligible amplitude of $A_{\rm TTV} = 1.102\times10^{-6}$ min. No coherent trend or periodicity is detected across the TESS dataset, indicating that WASP-19 b's orbit is dynamically stable and that any potential perturbations are below the current detection threshold.
- Given the absence of measurable TTVs, we explored whether a perturber capable of inducing the observed low-amplitude curvature could exist within dynamically stable regions. Using MEGNO maps and the `TTV2F2F` code (Hadden et al. 2019), we scanned the parameter space $(P'/P_b,\ M') \in ([0.1, 9.9], [10^{-6}, 10^{6}]\,M_\oplus)$ and found that all near-resonant solutions with $P' < 20$ d and $M' > 10\,M_\oplus$ are dynamically unstable or are ruled out by radial velocity limits. The remaining stable parameter space corresponds to perturbers below $\sim 10\,M_\oplus$ near the 1:2 and 2:1 resonances, but these would produce TTV amplitudes well below the current detection sensitivity.
- The observed curvature in the long-term dataset is best explained by a slow secular process, such as apsidal precession or weak tidal dissipation, rather than by an external perturber. Modeling the transit times with a precession term yields a rate of $\dot{\omega}_{\rm obs} = (1.00 \pm 0.12) \times 10^{-4}$ rad orbit$^{-1}$ and an associated Love number $k_{2p} = 0.107 \pm 0.08$, in agreement with Bernabò et al. (2024). The presence of the wide stellar companion WASP-19 B could further sustain the slight eccentricity required for ongoing precession through Kozai–Lidov oscillations.
- To assess robustness, we repeated the analysis on reduced subsets of the timing data after excluding the highest-uncertainty points (top 20–40%). In all cases, the cubic model remained marginally favored, but the absolute $\chi^2$ values remained $> 1$ and the residual scatter showed minimal improvement. This confirms that the statistical preference for a cubic fit arises primarily from the shape of the residual distribution rather than from genuine dynamical effects. The observed r.m.s. scatter of 46–50 s still exceeds the level required to detect secular orbital decay (~10 s over a decade), highlighting the need for future high-precision timing campaigns.
- Our photometric transmission spectrum of WASP-19 b shows no statistically significant evidence of TiO/VO absorption. After applying per-dataset offset corrections, the TiO/VO-free retrieval model yields a substantially higher Bayesian evidence ($\ln Z = 354.29 \pm 0.19$) and a lower reduced chi-square ($\chi^2_\nu = 1.85$). This improvement over our pre-offset analysis ($\chi^2_\nu = 2.74$) demonstrates that homogenizing the datasets leads to a more physically consistent interpretation. The preference for a TiO/VO-free atmosphere contrasts with previous high-resolution FORS2 detections (Sedaghati et al. 2017), a discrepancy most likely arising from the inherently lower spectral resolution of broadband photometry, which tends to smooth out narrow molecular features. If apsidal precession modulates irradiation over time, it could in principle lead to small-scale variability in the atmospheric temperature structure.

In summary, our analysis statistically favors the cubic ephemeris model for WASP-19 b, indicating measurable curvature in the long-term transit timing data. The most consistent physical interpretation of this trend is the apsidal precession of a slightly eccentric orbit, possibly sustained by long-term gravitational perturbations from the stellar companion WASP-19 B. Although the observed amplitude remains small relative to measurement uncertainties, the persistent preference for a non-linear model across independent datasets suggests that secular orbital evolution rather than tidal decay or stochastic noise is currently shaping the long-term behavior of WASP-19 b.

## 8. Discussion

The orbital and atmospheric properties of WASP-19 b have been extensively explored since its discovery, with numerous efforts focusing on detecting signs of orbital decay and probing its atmospheric composition (Hebb et al. 2010; Mancini et al. 2013; Sedaghati et al. 2017; Patra et al. 2020). Our comprehensive reanalysis of more than 15 years of transit observations reveals a statistically significant curvature in the transit timing residuals, favoring a *cubic* ephemeris model. This curvature, combined with the moderate period-change rate derived from our fits, provides evidence that WASP-19 b might be undergoing a long-term dynamical evolution dominated by apsidal precession rather than rapid tidal decay.

Our updated analysis yields a period change rate of $\dot{P} = -0.99 \pm 0.20$ ms yr$^{-1}$ (Table A.1), which places our result in close agreement with the value reported by Biswas et al. (2025, $\dot{P} = -1.1 \pm 0.4$ ms yr$^{-1}$) and within the same order of magnitude as the upper bound derived by Rosério & et al. (2022, $-1.40 \pm 0.88$ ms yr$^{-1}$). Our measured $\dot{P}$ is notably smaller than the earlier claim of rapid orbital decay reported by Patra et al. (2020, $\dot{P} = -6.50 \pm 1.33$ ms yr$^{-1}$) and slightly larger than the near-constant period solution of Ma et al. (2025). This intermediate value, together with the superior statistical performance of the cubic ephemeris, suggests that the orbital evolution of WASP-19 b is best explained by a slow secular process such as apsidal precession rather than monotonic tidal decay.

The cubic term encapsulates the non-linear curvature expected from precessional motion of a slightly eccentric orbit. The direct fitting of a precession model to the dataset yields a precession rate of $(1.00 \pm 0.12)\times10^{-4}$ rad orbit$^{-1}$ corresponding to a planetary Love number of $0.107 \pm 0.08$, which aligns closely with the value of $k_{2p} = 0.20^{+0.02}_{-0.03}$ reported by Bernabò et al. (2024). The consistency of these parameters across independent analyzes reinforces the interpretation that apsidal precession dominates the long-term timing evolution of WASP-19 b. The wide-separation stellar companion WASP-19 B may play a key dynamical role, inducing Kozai–Lidov oscillations or long-term secular forcing that sustain the small but nonzero eccentricity required for continued precession. Such secular interactions could also modulate the irradiation geometry of the planet over time, producing subtle variations in the structure and circulation of atmospheric temperature.

Although the cubic model is statistically favored (lowest AIC and BIC), the reduced values $\chi^2$ exceeding unity indicate that the timing dataset contains excess scatter beyond the formal uncertainties quoted. We recognize that this scatter is dominated by a subset of archival transit timings, particularly from older ground-based observations, for which the reported uncertainties may be underestimated. To mitigate this, we performed subset analyzes excluding high-uncertainty points (top 20–40%) and found that the cubic preference persisted, demonstrating that the inferred curvature is not driven by a small number of outliers or poorly constrained measurements. Moreover, the dedicated TESS-only analysis using `allesfitter` (Günther & Daylan 2019, 2021) and `TTVFast` (Deck et al. 2014) produced an excellent fit with negligible TTV amplitudes, indicating that short-period dynamical perturbations are not required to explain the data. This supports the interpretation that the long-term cubic trend reflects secular orbital evolution rather than stochastic noise or periodic forcing by an additional planet. Although we cannot entirely exclude the possibility of a low-mass external perturber, our MEGNO stability maps rule out companions more massive than $\sim 10\,M_\oplus$ near low-order resonances, thereby strengthening apsidal precession as the self-consistent physical explanation under the current observational constraints.

We note that occultation timings were intentionally excluded from this analysis due to their heterogeneous origins and instrument-dependent systematics. Restricting the analysis to homogeneously reduced transit timings ensures internal consistency over the 15-year baseline. Future observations combining precise secondary eclipse timings or phase-curve monitoring could further constrain eccentricity evolution and help discriminate between tidal and precessional contributions to the observed trend.

On the atmospheric side, the nature of the optical absorbers of WASP-19 b has long been debated. With the offset-corrected transmission spectrum, our retrieval analysis strongly favors a model without TiO/VO, yielding a higher Bayesian evidence and a lower reduced chi-square compared to the TiO/VO-included model. This statistically robust preference for a TiO/VO-free atmosphere contrasts with the tentative TiO detection reported by high-resolution VLT-FORS2 spectroscopy by Sedaghati et al. (2017). The discrepancy likely arises from fundamental differences in spectral resolution: broadband photometric transmission spectra smooth over narrow TiO/VO bandheads that high-resolution differential spectroscopy can isolate. Additionally, TiO and VO are expected to be much more abundant and radiatively active on the intensely irradiated dayside of hot-Jupiters, where temperatures are higher and thermal dissociation or cold-trapping are less efficient (Fortney et al. 2008; Parmentier et al. 2018). Because our analysis relies exclusively on transmission (terminator) measurements and does not include occultation (dayside emission) data, any TiO/VO present primarily on the dayside may remain undetectable in the limb spectrum. High-altitude haze and clouds at the terminator (Sing et al. 2016) may further obscure metal oxide features. Instead, our retrieval consistently identifies Na, K, $H_2O$, and Rayleigh scattering as the dominant opacity sources results consistent with earlier low-resolution photometric analyzes (Mancini et al. 2013) and physically plausible given the extreme irradiation of WASP-19 b's environment.

A critical aspect in transmission spectroscopy is the impact of stellar activity, as uncorrected stellar variability can introduce systematic biases in transit depth measurements. Starspots, faculae, and flares have been known to distort transit light curves, leading to incorrect interpretations of planetary spectral features (Ioannidis et al. 2016). Our study accounts for these effects by incorporating `PRISM+GEMC` starspot modeling, which has been used successfully in prior exoplanet studies (Tregloan-Reed et al. 2013; Mancini et al. 2013). By correcting for stellar contamination, our modeling ensures that either the transit timing measurements or the retrieved planetary transmission spectrum is not significantly biased by stellar inhomogeneities, reinforcing the robustness of our results.

## 9. Conclusions

Our study provides supporting evidence that the long-term orbital evolution of WASP-19 b could be dominated by the apsidal precession of a slightly eccentric orbit rather than rapid orbital decay. The statistically preferred cubic ephemeris, consistent precession rate, and derived Love number together point to a stable yet dynamically evolving system, possibly influenced by the distant stellar companion WASP-19 B. On the atmospheric side, our analysis supports a TiO/VO-free composition with prominent Na, K, and $H_2O$ features and potential Rayleigh scattering. By combining a dynamically consistent ephemeris model with robust retrievals corrected for stellar activity, this work presents a coherent picture of WASP-19 b as a hot-Jupiter whose orbital and atmospheric evolution are shaped primarily by long-term secular processes rather than by ongoing tidal decay. Looking ahead, the ability to distinguish between apsidal-precession and tidal-decay scenarios will critically depend on the availability of long-baseline high-quality transit timing data and on theoretical predictions of future transit times under both evolutionary pathways. While such forward modeling lies beyond the scope of the present work, the refined ephemeris and precession constraints reported here provide a robust empirical reference against which future, higher-precision measurements can be compared. As next-generation facilities such as JWST, PLATO, and the ELT achieve timing precisions at or below the ten-second level with improved control of systematics, WASP-19 b will become a benchmark system for probing long-term orbital evolution in hot-Jupiters and for assessing the role of data quality in detecting subtle secular effects.

*Acknowledgements.* We thank the anonymous referee and the editor for their constructive comments and valuable suggestions, which have significantly improved this manuscript. A.R.R acknowledges support from ANID Doctoral Scholarship program grant no.21231937. A.B acknowledges support from the Deutsche Forschungsgemeinschaft (DFG, German Research Foundation) under Germany's Excellence Strategy – EXC 2094 – 390783311. J.S acknowledges support from STFC under grant number ST/Y002563/1. F.A, M.A, and U.G.J acknowledge funding from the Novo Nordisk Foundation Interdisciplinary Synergy Programme grant no. NNF19OC0057374. T.C.H acknowledges funding from the Europlanet 2024 Research Infrastructure (RI) programme. The Europlanet 2024 RI provides free access to the world's largest collection of planetary simulation and analysis facilities, data services and tools, a ground-based observational network and programme of community support activities. Europlanet 2024 RI has received funding from the European Union's Horizon 2020 research and innovation programme under grant agreement No. 871149. R.F.J acknowledges the support provided by the GEMINI/ANID project under grant number 32240028, by ANID's Millennium Science Initiative through grant ICN12_009, awarded to the Millennium Institute of Astrophysics (MAS), and by ANID's Basal project FB210003. E.K is supported by the National Research Foundation of Korea 2021M3F7A1082056. The following internet-based resources were used in research for this paper: Exoplanet Transit Database (ETD); the SIMBAD database; NASA's Astrophysics Data System (ADS) Bibliograpic Services; We greatfully acknowledge the valuable suggestions and insightful discussions by Carlos Ferreria Lopez and Barbara Rojas, during the preparation of this manuscript.

## References

Albrecht, S., Winn, J. N., Johnson, J. A., et al. 2012, ApJ, 757, 18
Anderson, D. R., Smith, A. M. S., Madhusudhan, N., et al. 2013, MNRAS, 430, 3422
Antoniciello, G., Borsato, L., Lacedelli, G., et al. 2021, MNRAS, 505, 1567
Ballard, S., Fabrycky, D., Fressin, F., et al. 2011, ApJ, 743, 200
Barker, A. J., Braviner, H. J., & Ogilvie, G. I. 2016, MNRAS, 459, 924
Baştürk, Ö., Kutluay, A. C., Barker, A., et al. 2025, MNRAS, 541, 714
Bean, J. L., Désert, J.-M., Seifahrt, A., et al. 2013, ApJ, 771, 108
Bernabò, L. M., Csizmadia, S., Smith, A. M. S., et al. 2024, A&A, 684, A78
Birkby, J. L., Cappetta, M., Cruz, P., et al. 2014, MNRAS, 440, 1470
Biswas, S., Jiang, I.-G., Yeh, L.-C., et al. 2025, AJ, 170, 133
Bitsch, B., Lambrechts, M., & Johansen, A. 2015, A&A, 582, A112
Blackwell, D. E. & Shallis, M. J. 1977, MNRAS, 180, 177
Bonomo, A. S., Desidera, S., Benatti, S., et al. 2017, A&A, 602, A107
Borkovits, T., Érdi, B., Forgács-Dajka, E., & Kovács, T. 2003, A&A, 398, 1091
Boss, A. P. 1997, Science, 276, 1836
Chatterjee, S., Ford, E. B., Matsumura, S., & Rasio, F. A. 2008, ApJ, 686, 580
Cincotta, P. M. & Simó, C. 2000, A&AS, 147, 205
Cortés-Zuleta, P., Rojo, P., Wang, S., et al. 2020, A&A, 636, A98
Dawson, R. I. 2014, The Astrophysical Journal Letters, 790, L31
Deck, K. M., Agol, E., Holman, M. J., & Nesvorný, D. 2014, ApJ, 787, 132
Doyle, A. P., Davies, G. R., Smalley, B., Chaplin, W. J., & Elsworth, Y. 2014, MNRAS, 444, 3592
Dragomir, D., Kane, S. R., Pilyavsky, G., et al. 2011, AJ, 142, 115
Espinoza, N., Rackham, B. V., Jordán, A., et al. 2019, MNRAS, 482, 2065
Foreman-Mackey, D. 2016, The Journal of Open Source Software, 1, 24
Foreman-Mackey, D., Hogg, D. W., Lang, D., & Goodman, J. 2013, Publications of the Astronomical Society of the Pacific, 125, 306
Fortney, J. J., Lodders, K., Marley, M. S., & Freedman, R. S. 2008, ApJ, 678, 1419
Gallet, F., Bolmont, E., Mathis, S., Charbonnel, C., & Amard, L. 2017, A&A, 604, A112
Gomes, D. S., Ferreira, J. M., Ferreira, F., & Correia, A. C. M. 2021, A&A, 650, A178
Goswamy, T., Collier Cameron, A., & Wilson, T. G. 2024, MNRAS, 534, 843

Goździewski, K., Bois, E., Maciejewski, A. J., & Kiseleva-Eggleton, L. 2001, A&A, 378, 569
Gu, P.-G., Lin, D. N. C., & Bodenheimer, P. H. 2003, The Astrophysical Journal, 588, 509
Günther, M. N. & Daylan, T. 2019, Allesfitter: Flexible Star and Exoplanet Inference From Photometry and Radial Velocity, Astrophysics Source Code Library
Günther, M. N. & Daylan, T. 2021, ApJS, 254, 13
Hadden, S., Barclay, T., Payne, M. J., & Holman, M. J. 2019, AJ, 158, 146
Hadden, S. & Lithwick, Y. 2016, ApJ, 828, 44
Hadden, S. & Lithwick, Y. 2017, AJ, 154, 5
Haisch, K. E. J., Lada, E. A., & Lada, C. J. 2001, ApJL, 553, L153
Hebb, L., Collier-Cameron, A., Triaud, A. H. M. J., et al. 2010, ApJ, 708, 224
Hellier, C., Anderson, D. R., Collier-Cameron, A., et al. 2011, ApJ, 730, L31
Hinse, T. C., Christou, A. A., Alvarellos, J. L. A., & Goździewski, K. 2010, MNRAS, 404, 837
Holczer, T., Mazeh, T., Nachmani, G., et al. 2016, ApJS, 225, 9
Hou, Q. & Wei, X. 2022, MNRAS, 511, 3133
Hughes, I. G. & Hase, T. P. A. 2010, Measurements and their Uncertainties (New York: Oxford University Press)
Huitson, C. M., Sing, D. K., Pont, F., et al. 2013, MNRAS, 434, 3252
Ioannidis, P., Huber, K. F., & Schmitt, J. H. M. M. 2016, A&A, 585, A72
Jenkins, J. M., Twicken, J. D., McCauliff, S., et al. 2016, in Software and Cyberinfrastructure for Astronomy IV, Vol. 9913, SPIE, 1232–1251
Komacek, T. D. & Showman, A. P. 2016, ApJ, 821, 16
Komacek, T. D. & Showman, A. P. 2020, The Astrophysical Journal, 888, 2
Lambrechts, M. & Johansen, A. 2014, A&A, 572, A107
Langton, J. & Laughlin, G. 2008, The Astrophysical Journal, 674, 1106–1116
Lendl, M., Gillon, M., Queloz, D., et al. 2013, A&A, 552, A2
Levrard, B., Winisdoerffer, C., & Chabrier, G. 2009, ApJ, 692, L9
Lomb, N. R. 1976, Ap&SS, 39, 447
Love, H. 1912, nature, 89, 1476
Ma, X., Wang, W., Zhang, Z., et al. 2025, AJ, 169, 169
MacDonald, R. J. 2023, The Journal of Open Source Software, 8, 4873
MacDonald, R. J. & Madhusudhan, N. 2017, MNRAS, 469, 1979
Maciejewski, G. 2019, Contributions of the Astronomical Observatory Skalnate Pleso, 49, 334
Maciejewski, G., Fernández, M., Aceituno, F., et al. 2018, Acta Astron., 68, 371
Maggio, A., Sanz-Forcada, J., Scandariato, G., & et al. 2012, A&A, 544, A23
Mancini, L., Ciceri, S., Chen, G., et al. 2013, MNRAS, 436, 2
Mandell, A. M., Haynes, K., Sinukoff, E., et al. 2013, ApJ, 779, 128
Mathis, S. 2019, in EAS Publications Series, Vol. 82, EAS Publications Series, 5–33
Matsumura, S., Peale, S. J., & Rasio, F. A. 2010, ApJ, 725, 1995
May, E. J. & Rauscher, E. 2021, The Astronomical Journal, 161, 159
Mayor, M. & Queloz, D. 1995, Nature, 378, 355
Mordasini, C. 2018, Handbook of Exoplanets
Morrell, S., Naylor, T., Southworth, J., & Sing, D. K. 2026, MNRAS, 545, staf2063
Ogilvie, G. I. 2014, ARA&A, 52, 171
Ogilvie, G. I. & Lin, D. N. C. 2004, The Astrophysical Journal, 610, 477
Oshagh, M., Santos, N. C., Ehrenreich, D., et al. 2014, Astronomy and Astrophysics, 568, A99
Paardekooper, S., Dong, R., Duffell, P., et al. 2023, in Astronomical Society of the Pacific Conference Series, Vol. 534, Protostars and Planets VII, ed. S. Inutsuka, Y. Aikawa, T. Muto, K. Tomida, & M. Tamura, 685
Parmentier, V., Line, M. R., Bean, J. L., et al. 2018, A&A, 617, A110
Patra, K. C., Winn, J. N., Holman, M. J., et al. 2020, The Astronomical Journal, 159, 150
Patra, K. C., Winn, J. N., Holman, M. J., et al. 2017, AJ, 154, 4
Penev, K., Bouma, L. G., Winn, J. N., & Hartman, J. D. 2018, AJ, 155, 165
Petrucci, R., Jofré, E., Gómez Maqueo Chew, Y., et al. 2020, MNRAS, 491, 1243
Pollack, J. B., Hubickyj, O., Bodenheimer, P., et al. 1996, Icarus, 124, 62
Pont, F., Knutson, H., Gilliland, R. L., Moutou, C., & Charbonneau, D. 2008, Monthly Notices of the Royal Astronomical Society, 385, 109–118
Rackham, B. V., Apai, D., & Giampapa, M. S. 2018, The Astrophysical Journal, 853, 122
Ragozzine, D. & Wolf, A. S. 2009, ApJ, 698, 1778
Rasio, F. A. & Ford, E. B. 1996, Science, 274, 954
Rauscher, E. & Kempton, E. M. R. 2014, The Astrophysical Journal, 790, 79
Ricker, G. R., Winn, J. N., Vanderspek, R., et al. 2015, Journal of Astronomical Telescopes, Instruments, and Systems, 1, 014003
Rosério, C. & et al. 2022, Astronomy & Astrophysics, 664, A89
Sanchis-Ojeda, R., Winn, J. N., Holman, M. J., et al. 2011a, ApJ, 733, 127
Sanchis-Ojeda, R., Winn, J. N., Holman, M. J., et al. 2011b, ApJ, 733, 127
Scargle, J. D. 1982, ApJ, 263, 835
Sedaghati, E., Boffin, H. M. J., Csizmadia, S., et al. 2015a, A&A, 576, L11
Sedaghati, E., Boffin, H. M. J., Csizmadia, S., et al. 2015b, A&A, 576, L11
Sedaghati, E., Boffin, H. M. J., MacDonald, R. J., et al. 2017, Nature, 549, 238
Sing, D. K., Fortney, J. J., Nikolov, N., et al. 2016, Nature, 529, 59
Southworth, J., Hinse, T. C., Burgdorf, M., et al. 2014, MNRAS, 444, 776
Southworth, J., Hinse, T. C., Burgdorf, M. J., et al. 2009a, MNRAS, 399, 287
Southworth, J., Hinse, T. C., Jørgensen, U. G., et al. 2009b, MNRAS, 396, 1023
Stetson, P. B. 1987, PASP, 99, 191
Stumpe, M. C., Smith, J. C., Catanzarite, J. H., et al. 2014, Publications of the Astronomical Society of the Pacific, 126, 100
Tregloan-Reed, J., Southworth, J., Burgdorf, M., et al. 2015, MNRAS, 450, 1760
Tregloan-Reed, J., Southworth, J., Mancini, L., et al. 2018, MNRAS, 474, 5485
Tregloan-Reed, J., Southworth, J., & Tappert, C. 2013, MNRAS, 428, 3671
Tregloan-Reed, J. & Unda-Sanzana, E. 2019, A&A, 630, A114
Tregloan-Reed, J. & Unda-Sanzana, E. 2021, A&A, 649, A130
Valsecchi, F. & Rasio, F. A. 2014, ApJ, 787, L9
VanderPlas, J. T. 2018, ApJS, 236, 16
Watson, C. A. & Marsh, T. R. 2010, MNRAS, 405, 2037
Wong, I., Knutson, H. A., Kataria, T., et al. 2016, ApJ, 823, 122
Yahalomi, D. A. & Kipping, D. 2026, ApJ, 998, 136
Yalçınkaya, S., Şenavcı, H. V., & Özavcı, 2024, MNRAS, 530, 2475–2487

[1] Instituto de Astronomia y Ciencias Planetarias de Atacama (INCT), Universidad de Atacama, Copayapu 485, Copiapo, Chile e-mail: anitha.raj.21@alumnos.uda.cl
[2] European Southern Observatory, Karl-Schwarzschild-Strasse 2, 85748 Garching bei Munchen, Germany
[3] Department of Physics, Washington University, St. Louis, MO 63130, USA
[4] School of Earth and Space Sciences, University of Science and Technology of China, Hefei 230026, China
[5] Centro de Astronomia (CITEVA), Universidad de Antofagasta, Antofagasta, Chile
[6] Astrophysics Group, Keele University, Staffordshire ST5 5BG, UK
[7] University of Southern Denmark, Department of Physics, Chemistry and Pharmacy, Odense, Denmark
[8] Centre for Astrophysics Research, University of Hertfordshire, UK
[9] Institute for Astronomy, University of Edinburgh, UK

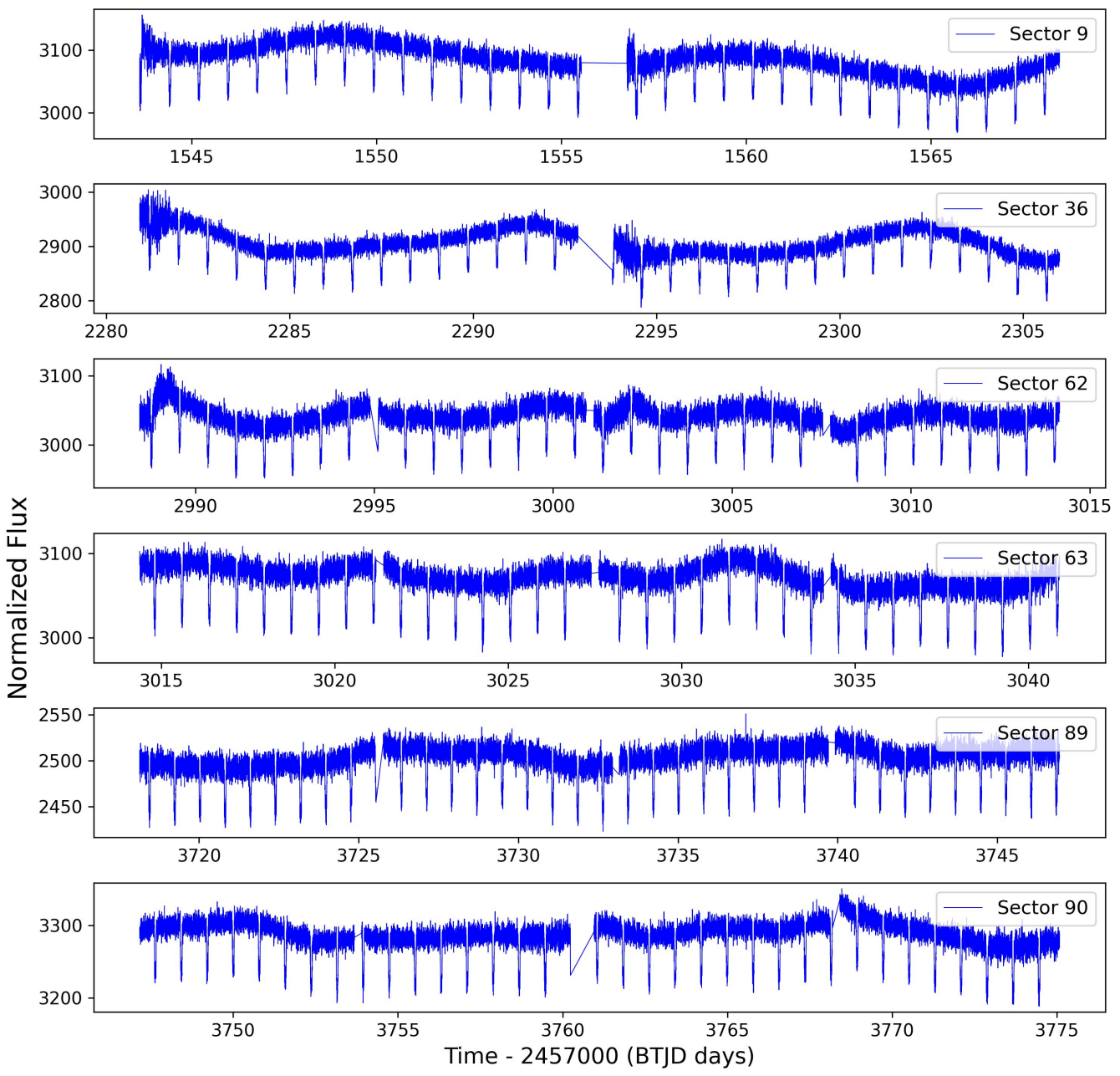

**Fig. A.1.** TESS photometric lightcurves of the Sector 9, 36, 62, 63, 89 and 90

## Appendix A: Robustness tests for ephemeris models

This presents a set of robustness tests designed to evaluate whether the detected departure from strictly steady timing variations depends on the inclusion of specific sub-datasets, particularly the lower precision but temporally extensive TESS observations. Table A.3 compares the statistical performance of linear, quadratic, and cubic ephemeris models for datasets constructed with and without TESS timings.

TESS, with its space-based precision, provides extensive temporal coverage that is invaluable for long-term orbital analysis. However, its smaller aperture compared to ground-based telescopes like the Danish 1.54-m results in higher scatter in photometric measurements. The motivation for this comparative experiment is therefore not to reassess the dataset itself, but to determine whether the preference for the cubic ephemeris persists when the noisier TESS subset and outliers are removed.

Key observations and results: From the combined analysis from Table A.3 and A.4, the following conclusions include:

- Across all data-quality cuts (full sample, best 20–40 % timing precision, and removal of high-uncertainty or large-residual points), the cubic ephemeris consistently yields the lowest AIC and BIC values. This indicates that the increased model complexity is statistically supported rather than an artifact of any specific subset.
- The inclusion of TESS data increases the overall $\chi^2$ and RMS residuals as expected from its higher photometric scatter but importantly, it does not erase or reverse the statistical preference for the cubic model. Both AIC and BIC continue to favor the cubic form in the "With TESS" and "Without TESS" cases.
- For the highest-precision subsets (best 20–40 %), the cubic model maintains substantially lower information-criterion values, demonstrating that the non-linear curvature in the timing residuals is present even when the lowest-precision timestamps are removed.
- To further test whether the cubic trend could be artificially driven by a few influential points, we also conducted a large-scale random hold-out cross-validation ($X_{\rm val}$). In this test, 20 % of the transit times were randomly removed and the remaining 80 % refitted, repeated over $10^5$ independent iterations. For each iteration, we compared the prediction errors of the linear, quadratic, and cubic ephemerides on the held-out data. Across all iterations, the cubic model produced prediction errors with a median absolute deviation comparable to the simpler models, while exhibiting the least systematic bias, indicating that the global curvature persists regardless of which measurements are withheld.

**Table A.1.** Statistical parameters for different subsets of the transit timing dataset.

| Subsets | $Q'_*$ | $dP/dE$ | $dP/dt$ (m s yr$^{-1}$) |
|---|---|---|---|
| **All data without TESS** | $10^{6.16}$ | -5.319517e-11 | -1.348866 |
| Best 20% without TESS | $10^{6.20}$ | -4.902598e-11 | -1.243149 |
| Best 30% without TESS | $10^{6.167}$ | -5.285915e-11 | -1.340346 |
| Best 40% without TESS | $10^{6.34}$ | -3.511581e-11 | -0.890429 |
| **All data with TESS** | $10^{6.29}$ | -3.924766e-11 | -0.995200 |
| Best 20% with TESS | $10^{6.58}$ | -2.025481e-11 | -0.513600 |
| Best 30% with TESS | $10^{6.48}$ | -2.526531e-11 | -0.640651 |
| Best 40% with TESS | $10^{6.66}$ | -1.700657e-11 | -0.431235 |

**Notes.** Results are presented with and without the inclusion of TESS data, using the full transit sample as well as the best 14%, 21%, and 40% subsets of the dataset.

In summary, this comparative analysis demonstrates that the detected departure from strictly steady timing variations is recovered regardless of whether TESS data are included. Although the TESS timings introduce additional scatter, both the AIC and BIC consistently favor the cubic ephemeris over the linear and quadratic forms, indicating that the increased model complexity is statistically warranted rather than an artifact of the lower-precision subset. Cross-validation experiments further confirm that the inferred curvature is not driven by isolated outliers or by any single sub-dataset. These results highlight the robustness of the non-linear trend detected in the timing residuals and underline the importance of continued, high-precision ground- and space-based monitoring to refine the long-term dynamical interpretation of the WASP-19 system.

## Appendix B: TTV analysis

This focused TTV analysis aims to test the possibility that the transit timing residuals observed in Fig. 3 are due to dynamical interactions with an unseen perturber in the WASP-19 system. To measure TTVs for TESS mid-transit times, we used open-source software `allesfitter` (Günther & Daylan 2019, 2021) that models the time-series TESS photometry data by performing MCMC built from the `emcee` (Foreman-Mackey et al. 2013) ensemble sampler and the `corner` (Foreman-Mackey 2016) package. We set the priors to run 100,000 steps with 200 walkers and 6,000 burn-in steps to ensure a reasonable sampling process. After exploring all correlations in the parameter space, the TTVs and other orbital parameters can be accurately measured with their uncertainties estimated by the `allesfitter` software.

Next, we applied the N-body simulation `TTVFast` (Deck et al. 2014) to evaluate the long-term TTV series. For a planetary system with a known transiting planet and a potential unseen perturber, this synthetic TTV series can be determined by

**Table A.2.** Mid-transit timing and photometric parameters for WASP-19 b used in this work.

| Date | $BJD_{TDB}$ | $\sigma$T (s) | Cycle | Resi -dual (s) | (O-C)/ $\sigma$T | $r_p/r_s$ | $r_s + r_p$ | $i$ | Filter | Source |
|---|---|---|---|---|---|---|---|---|---|---|
| Nov 04, 2008 | 2454775.3372 | ±13 | -517.0 | -55 | -4.23 | $0.1424 \pm 0.02$ | $0.3222 \pm 0.03$ | 80.5±0.7 | $V + R$ | 1 |
| Dec 03, 2009 | 2455168.968 | ±10 | -18.0 | -44 | -4.40 | $0.1425 \pm 0.0014$ | $0.3432 \pm 0.002$ | $78.02 \pm 2$ | V | 2 |
| Dec 17, 2009 | 2455183.16711 | ±6 | 0.0 | -45 | -7.50 | $0.1431 \pm 0.0015$ | $0.3356 \pm 0.0032$ | $79.42 \pm 0.39$ | Gunn $r$ | 3 |
| Feb 24, 2010 | 2455251.7962 | ±12 | 87.0 | -37 | -3.08 | 0.1435±0.0014 | 0.3298±0.0041 | 78.97±0.39 | Gunn $r$ | 4 |
| Feb 25, 2010 | 2455252.5851 | ±9 | 88.0 | -35 | -3.89 | 0.1417±0.0013 | 0.3300±0.0025 | 78.92±0.37 | Gunn $r$ | 4 |
| Feb 28, 2010 | 2455255.7404 | ±10 | 92.0 | -37 | -3.70 | 0.1430±0.0008 | 0.3311±0.0044 | 78.91±0.44 | Gunn $r$ | 4 |
| Feb 28, 2010 | 2455255.74116 | ±8 | 92.0 | 29 | 3.63 | $0.1433^{+0.0018}_{-0.0030}$ | $0.3282^{+0.0070}_{-0.0063}$ | $78.85^{+0.55}_{-0.66}$ | Gunn-r | 5 |
| Mar 04, 2010 | 2455259.6846 | ±27 | 97.0 | -37 | -1.37 | $0.1405^{+0.0048}_{-0.0139}$ | $0.2999^{+0.0336}_{-0.0313}$ | $81.38^{+4.30}_{-2.99}$ | clear | 6 |
| Mar 18, 2010 | 2455273.8825 | ±54 | 115.0 | -138 | -2.56 | $0.1416^{+0.0081}_{-0.0077}$ | $0.3257^{+0.0502}_{-0.0493}$ | $80.43^{+5.06}_{-4.52}$ | clear | 6 |
| May 17, 2010 | 2455334.6255 | ±15 | 192.0 | 67 | 4.47 | 0.1397±0.0018 | 0.3170±0.0091 | 80.15±0.80 | R | 7 |
| May 21, 2010 | 2455338.5688 | ±18 | 197.0 | -8 | -0.44 | 0.1435±0.0017 | 0.3328±0.0107 | 78.48±0.84 | I+z′ | 8 |
| Jun 05, 2010 | 2455353.9514 | ±21 | 216.5 | 9 | 0.43 | 0.1390±0.0023 | 0.3128±0.0117 | 80.82±1.12 | R | 7 |
| Jun 20, 2010 | 2455368.5446 | ±30 | 235.0 | -21 | -0.70 | 0.1478±0.0034 | 0.3442±0.0170 | 79.00±1.43 | R | 7 |
| Dec 09, 2010 | 2455539.7233 | ±20 | 452.0 | 35 | 1.75 | 0.1379±0.0036 | 0.3199±0.0200 | 80.28±1.98 | IC | 8 |
| Jan 08, 2011 | 2455569.6976 | ±18 | 490.0 | -105 | -5.83 | 0.1397±0.0027 | 0.3254±0.0134 | 79.83±1.28 | I+z′ | 8 |
| Jan 19, 2011 | 2455580.742 | ±50 | 504.0 | -44 | -0.88 | $0.1431 \pm 0.013$ | $0.3827 \pm 0.021$ | $87.2 \pm 2.9$ | Cousins $R$ | 9 |
| Jan 23, 2011 | 2455584.6867 | ±15 | 509.0 | 1 | 0.07 | 0.1445±0.0019 | 0.3387±0.0100 | 78.26±0.89 | r′ | 8 |
| Jan 23, 2011 | 2455602.8303 | ±18 | 532.0 | 20 | 1.11 | 0.1473±0.0021 | 0.3504±0.0120 | 77.19±1.00 | I+z′ | 8 |
| Feb 01, 2011 | 2455594.1523 | ±43 | 521.0 | -39 | -0.91 | $0.1365^{+0.0081}_{-0.0074}$ | $0.2759^{+0.0576}_{-0.0268}$ | $83.53^{+6.33}_{-3.62}$ | $R_C$ | 6 |
| Feb 08, 2011 | 2455601.2524 | ±30 | 530.0 | 1 | 0.03 | $0.1471^{+0.0054}_{-0.0167}$ | $0.3501^{+0.0361}_{-0.0399}$ | $77.37^{+3.51}_{-2.90}$ | $R_C$ | 6 |
| Feb 14, 2011 | 2455606.7742 | ±17 | 537.0 | 0 | - 0.00 | 0.1443±0.0021 | 0.0120±0.3328 | 79.08±0.98 | z′ | 8 |
| Feb 15, 2011 | 2455607.5615 | ±29 | 538.0 | -132 | -4.55 | 0.1394±0.0048 | 0.3040±0.0224 | 81.56±2.55 | I+z′ | 8 |
| Mar 03, 2011 | 2455624.1284 | ±41 | 559.0 | -21 | -0.51 | $0.1329^{+0.0082}_{-0.0167}$ | $0.3230^{+0.0698}_{-0.0415}$ | $79.43^{+6.75}_{-5.21}$ | $R_C$ | 6 |
| Apr 04, 2011 | 2455655.6814 | ±20 | 599.0 | -69 | -3.45 | 0.1336±0.0035 | 0.2971±0.0109 | 80.62±0.87 | I+z′ | 8 |
| May 07, 2011 | 2455689.6026 | ±27 | 642.0 | 26 | 0.96 | 0.1432±0.0083 | 0.3405±0.0230 | 78.33±1.88 | R | 7 |
| May 11, 2011 | 2455693.9406 | ±60 | 647.5 | -28 | -0.47 | 0.1376±0.0011 | 0.3276±0.0082 | 79.26±0.65 | R | 7 |
| May 22, 2011 | 2455704.5902 | ±24 | 661.0 | -2 | -0.08 | 0.1379±0.0024 | 0.3539±0.0166 | 76.91±1.39 | R | 7 |
| May 26, 2011 | 2455708.5349 | ±16 | 666.0 | 35 | 2.19 | 0.1353±0.0027 | 0.2981±0.0133 | 82.21±1.32 | R | 7 |
| Mar 13, 2012 | 2455999.6163 | ±16 | 1035.0 | 24 | 1.50 | $0.1407^{+0.0014}_{-0.0024}$ | $0.3274^{+0.0142}_{-0.0114}$ | $78.66^{+0.96}_{-1.18}$ | 1.25-1.33 $\mu$m | 10 |
| Mar 13, 2012 | 2455999.6165 | ±15 | 1035.0 | 37 | 2.47 | $0.1450^{+0.0025}_{-0.0032}$ | $0.3282^{+0.0120}_{-0.0139}$ | $79.11^{+1.26}_{-1.00}$ | 1.4-1.5 $\mu$m | 10 |
| Mar 13, 2012 | 2455999.6164 | ±10 | 1035.0 | 25 | 2.50 | $0.1564^{+0.0012}_{-0.0022}$ | $0.3196^{+0.0038}_{-0.0064}$ | $79.70^{+0.56}_{-0.52}$ | 1.5-1.6 $\mu$m | 10 |
| Mar 13, 2012 | 2455999.6160 | ±15 | 1035.0 | -1 | -0.07 | $0.1434^{+0.0022}_{-0.0025}$ | $0.3271^{+0.0079}_{-0.0095}$ | $78.82^{+0.83}_{-0.68}$ | 1.6-1.7 $\mu$m | 10 |
| Mar 13, 2012 | 2455999.6161 | ±22 | 1035.0 | 6 | 0.27 | $0.1376^{+0.0029}_{-0.0031}$ | $0.3207^{+0.0103}_{-0.0101}$ | $79.40^{+0.71}_{-0.74}$ | 1.7-1.8 $\mu$m | 10 |
| Mar 13, 2012 | 2455999.6163 | ±22 | 1035.0 | 23 | 1.05 | $0.1436^{+0.0015}_{-0.0016}$ | $0.3246^{+0.0106}_{-0.0105}$ | $78.90^{+0.91}_{-0.91}$ | 1.95-2.05 $\mu$m | 10 |
| Mar 13, 2012 | 2455999.6164 | ±10 | 1035.0 | 26 | 2.60 | $0.1441^{+0.0011}_{-0.0011}$ | $0.3377^{+0.0070}_{-0.0075}$ | $77.93^{+0.69}_{-0.60}$ | 2.05-2.15 $\mu$m | 10 |
| Mar 13, 2012 | 2455999.6165 | ±16 | 1035.0 | 38 | 2.38 | $0.1413^{+0.0012}_{-0.0024}$ | $0.3123^{+0.0088}_{-0.0105}$ | $79.95^{+0.96}_{-0.80}$ | 2.15-2.25 $\mu$m | 10 |
| Mar 13, 2012 | 2455999.6168 | ±25 | 1035.0 | 59 | 2.36 | $0.1375^{+0.0030}_{-0.0041}$ | $0.2917^{+0.0242}_{-0.0203}$ | $82.07^{+2.53}_{-2.24}$ | 2.25-2.35 $\mu$m | 10 |
| Apr 04, 2012 | 2456021.7036 | ±11 | 1063.0 | 2 | 0.18 | $0.1484^{+0.0014}_{-0.0018}$ | $0.3532^{+0.0073}_{-0.0093}$ | $76.70^{+0.79}_{-0.68}$ | 1.25-1.33 $\mu$m | 10 |
| Apr 04, 2012 | 2456021.7033 | ±12 | 1063.0 | -19 | -1.58 | $0.1473^{+0.0013}_{-0.0014}$ | $0.3347^{+0.0093}_{-0.0093}$ | $78.28^{+0.82}_{-0.79}$ | 1.4-1.5 $\mu$m | 10 |
| Apr 04, 2012 | 2456021.7036 | ±9 | 1063.0 | 6 | 0.67 | $0.1461^{+0.0013}_{-0.0016}$ | $0.3320^{+0.0121}_{-0.0075}$ | $78.59^{+0.70}_{-1.07}$ | 1.5-1.6 $\mu$m | 10 |
| Apr 04, 2012 | 2456021.7041 | ±9 | 1063.0 | 45 | 5.00 | $0.1453^{+0.0013}_{-0.0025}$ | $0.3293^{+0.0067}_{-0.0069}$ | $78.77^{+0.70}_{-0.60}$ | 1.6-1.7 $\mu$m | 10 |
| Apr 04, 2012 | 2456021.7038 | ±10 | 1063.0 | 21 | 2.10 | $0.1437^{+0.0012}_{-0.0014}$ | $0.3379^{+0.0068}_{-0.0069}$ | $78.04^{+0.54}_{-0.63}$ | 1.7-1.8 $\mu$m | 10 |
| Apr 04, 2012 | 2456021.7032 | ±13 | 1063.0 | -29 | -2.23 | $0.1513^{+0.0012}_{-0.0019}$ | $0.3279^{+0.0080}_{-0.0102}$ | $79.08^{+0.86}_{-0.68}$ | 1.95-2.05 $\mu$m | 10 |
| Apr 04, 2012 | 2456021.7039 | ±9 | 1063.0 | 32 | 3.56 | $0.1441^{+0.0014}_{-0.0017}$ | $0.3353^{+0.0077}_{-0.0074}$ | $77.93^{+0.70}_{-0.60}$ | 2.05-2.15 $\mu$m | 10 |
| Apr 04, 2012 | 2456021.7037 | ±10 | 1063.0 | 13 | 1.30 | $0.1406^{+0.0011}_{-0.0017}$ | $0.3217^{+0.0089}_{-0.0122}$ | $79.29^{+1.09}_{-0.80}$ | 2.15-2.25 $\mu$m | 10 |
| Apr 04, 2012 | 2456021.7043 | ±36 | 1063.0 | 63 | 1.75 | $0.1434^{+0.0045}_{-0.0040}$ | $0.3196^{+0.0137}_{-0.0159}$ | $79.51^{+1.34}_{-1.01}$ | 2.25-2.35 $\mu$m | 10 |
| Apr 15, 2012 | 2456033.5363 | ±12 | 1078.0 | 11 | 0.92 | 0.1434±0.0015 | 0.3318±0.0072 | 78.64±0.66 | Gunn $i'$ | 7 |
| Apr 15, 2012 | 2456033.5365 | ±11 | 1078.0 | 32 | 2.91 | 0.1414±0.0014 | 0.3333±0.0066 | 78.70±0.54 | Gunn $z'$ | 7 |
| Apr 15, 2012 | 2456033.5362 | ±12 | 1078.0 | 9 | 0.75 | 0.1417±0.1426 | 0.3296±0.3262 | 79.12±0.76 | Gunn $r'$ | 7 |
| Apr 15, 2012 | 2456033.5362 | ±12 | 1078.0 | 2 | 0.17 | 0.1417±0.0019 | 0.3291±0.0082 | 78.63±0.71 | Gunn $g'$ | 7 |
| Apr 15, 2012 | 2456033.5373 | ±14 | 1078.0 | 102 | 7.29 | 0.1309±0.0018 | 0.3247±0.0090 | 78.89±0.79 | J | 7 |
| Apr 15, 2012 | 2456033.5376 | ±21 | 1078.0 | 126 | 6.00 | 0.1388±0.0043 | 0.3270±0.0180 | 79.94±1.58 | H | 7 |
| Apr 15, 2012 | 2456033.5366 | ±21 | 1078.0 | 42 | 2.00 | 0.1299±0.0053 | 0.3141±0.0209 | 79.86±2.09 | K | 7 |
| May 15, 2012 | 2456063.9054 | ±25 | 1116.5 | -91 | -3.64 | 0.1451±0.0027 | 0.3625±0.0164 | 76.25±1.29 | I+z′ | 8 |
| Feb 11, 2013 | 2456334.8721 | ±39 | 1460.0 | -52 | -1.33 | $0.1408^{+0.0107}_{-0.0098}$ | $0.3203^{+0.0377}_{-0.0295}$ | $79.08^{+2.98}_{-3.00}$ | clear | 6 |

**Table A.2.** Continued.

| Date | $BJD_{TDB}$ | $\sigma$T (s) | Cycle | Residual (s) | (O-C)/$\sigma$T | $r_p/r_s$ | $r_s + r_p$ | $i$ | Filter | Source |
|---|---|---|---|---|---|---|---|---|---|---|
| Apr 08, 2013 | 2456390.8803 | ±40 | 1531.0 | 7 | 0.18 | $0.1365^{+0.0064}_{-0.0097}$ | $0.3110^{+0.0327}_{-0.0282}$ | $80.26^{+2.05}_{-2.61}$ | clear | 6 |
| May 24, 2013 | 2456436.6329 | ±48 | 1589.0 | -3 | -0.06 | $0.1419^{+0.0094}_{-0.0081}$ | $0.3002^{+0.0353}_{-0.0453}$ | $81.65^{+7.14}_{-2.96}$ | clear | 6 |
| Mar 03, 2014 | 2456719.8262 | ±41 | 1948.0 | 8 | 0.20 | $0.1287^{+0.0071}_{-0.0179}$ | $0.2884^{+0.0698}_{-0.0325}$ | $82.01^{+7.79}_{-4.98}$ | clear | 6 |
| Mar 22, 2014 | 2456739.5477 | ±11 | 1973.0 | 47 | 4.27 | $0.1361^{+0.0045}_{-0.0060}$ | $0.3207^{+0.0239}_{-0.0159}$ | $79.24^{+0.94}_{-1.76}$ | clear | 11 |
| Mar 23, 2014 | 2456739.5474 | ±22 | 1973.0 | 21 | 0.95 | $0.1425^{+0.0026}_{-0.0104}$ | $0.3364^{+0.0127}_{-0.0253}$ | $77.99^{+2.27}_{-1.17}$ | clear | 6 |
| Nov 15, 2014 | 2456977.7765 | ±8 | 2275.0 | -3 | -0.38 | $0.1390^{+0.0014}_{-0.0016}$ | $0.3176^{+0.0021}_{-0.0022}$ | $80.06^{+0.19}_{-0.20}$ | clear | 12 |
| Apr 25, 2014 | 2456772.6779 | ±45 | 2015.0 | -40 | -0.89 | $0.1410 \pm 0.015$ | $0.3215 \pm 0.017$ | $81.23 \pm 2.12$ | $I + z'$ | 13 |
| Nov 16, 2014 | 2456977.7772 | ±11 | 2275.0 | 62 | 5.64 | $0.1416^{+0.0019}_{-0.0018}$ | $0.3214^{+0.0023}_{-0.0024}$ | $80.36^{+0.76}_{-0.81}$ | GG435 | 14 |
| May 03, 2015 | 2457146.5880 | ±10 | 2489.0 | -3 | -0.30 | $0.1437 \pm 0.0011$ | $0.3210 \pm 0.0056$ | $80.01 \pm 0.51$ | R | 15 |
| May 07, 2015 | 2457150.5320 | ±11 | 2494.0 | -24 | -2.18 | $0.1404 \pm 0.0017$ | $0.3277 \pm 0.0070$ | $79.67 \pm 0.60$ | R | 15 |
| May 18, 2015 | 2457161.5754 | ±25 | 2508.0 | -51 | -2.04 | $0.1440 \pm 0.0013$ | $0.3323 \pm 0.0042$ | $78.95 \pm 0.37$ | R | 15 |
| May 22, 2015 | 2457165.5203 | ±35 | 2513.0 | 10 | 0.29 | $0.1402 \pm 0.0054$ | $0.3310 \pm 0.0290$ | $79.09 \pm 2.38$ | R | 15 |
| Jun 02, 2015 | 2457176.5638 | ±23 | 2527.0 | -12 | -0.52 | $0.1381 \pm 0.0036$ | $0.3211 \pm 0.0194$ | $79.84 \pm 1.83$ | R | 15 |
| Jun 06, 2015 | 2457180.5081 | ±15 | 2532.0 | -8 | -0.53 | $0.1439 \pm 0.0022$ | $0.0090 \pm 0.3323$ | $78.93 \pm 0.80$ | R | 15 |
| Jan 30, 2016 | 2457418.7368 | ±17 | 2834.0 | -63 | -3.71 | $0.1392^{+0.0018}_{-0.0022}$ | $0.3324^{+0.0048}_{-0.0052}$ | $78.33^{+0.47}_{-0.36}$ | clear | 12 |
| Feb 29, 2016 | 2457448.7129 | ±6 | 2872.0 | -44 | -7.33 | $0.1474^{+0.0017}_{-0.0018}$ | $0.3205^{+0.0038}_{-0.0030}$ | $79.46^{+0.31}_{-0.38}$ | clear | 12 |
| Apr 26, 2016 | 2457505.5098 | ±12 | 2944.0 | -8 | -0.67 | $0.1415 \pm 0.0017$ | $0.3248 \pm 0.0073$ | $79.34 \pm 0.63$ | R | 15 |
| Apr 29, 2016 | 2457508.6652 | ±12 | 2948.0 | 0 | 0.00 | $0.1430 \pm 0.0018$ | $0.3318 \pm 0.0078$ | $78.92 \pm 0.70$ | R | 15 |
| May 03, 2016 | 2457512.6098 | ±21 | 2953.0 | 34 | 1.62 | $0.1463 \pm 0.0029$ | $0.3555 \pm 0.0152$ | $77.69 \pm 1.35$ | R | 15 |
| May 07, 2016 | 2457516.5537 | ±13 | 2958.0 | 9 | 0.69 | $0.1469 \pm 0.0014$ | $0.3399 \pm 0.0070$ | $77.77 \pm 0.59$ | R | 15 |
| May 09, 2016 | 2457518.1318 | ±33 | 2960.0 | 46 | 1.39 | $0.1391^{+0.0041}_{0.0107}$ | $0.3167^{+0.0362}_{-0.0322}$ | $80.03^{+3.83}_{2.99}$ | $R_C$ | 16 |
| Jun 07, 2016 | 2457546.5303 | ±33 | 2996.0 | 67 | 2.03 | $0.1410 \pm 0.015$ | $0.3215 \pm 0.017$ | $81.23 \pm 2.12$ | $I + z'$ | 13 |
| Jun 06, 2016 | 2457546.5298 | ±24 | 2996.0 | 24 | 1.00 | $0.1442^{+0.0028}_{-0.0096}$ | $0.3208^{+0.0297}_{-0.0216}$ | $79.39^{+2.13}_{-2.66}$ | R | 16 |
| Jun 06, 2016 | 2457546.5297 | ±13 | 2996.0 | 23 | 1.77 | $0.1327 \pm 0.0037$ | $0.2835 \pm 0.0153$ | $84.37 \pm 1.92$ | R | 15 |
| Jun 10, 2016 | 2457550.4727 | ±39 | 3001.0 | -84 | -2.15 | $0.1472^{+0.0075}_{-0.0064}$ | $0.3234^{+0.0334}_{-0.0378}$ | $79.20^{+2.87}_{-2.30}$ | R | 16 |
| Jun 21, 2016 | 2457561.5179 | ±12 | 3015.0 | 42 | 3.50 | $0.1437 \pm 0.0020$ | $0.3258 \pm 0.0082$ | $79.23 \pm 0.76$ | R | 15 |
| Jul 06, 2016 | 2457576.5047 | ±64 | 3034.0 | -57 | -0.89 | $0.1356^{+0.0094}_{-0.0086}$ | $0.3130^{+0.0467}_{-0.0506}$ | $80.52^{+5.97}_{-3.58}$ | R | 16 |
| Dec 02, 2016 | 2457724.8078 | ±62 | 3222.0 | 62 | 1.00 | $0.1616^{+0.0299}_{-0.0196}$ | $0.3300^{+0.0242}_{-0.0173}$ | $78.86^{+1.19}_{-1.02}$ | I | 6 |
| Dec 16, 2016 | 2457739.7942 | ±47 | 3241.0 | -77 | -1.64 | $0.1521^{+0.0144}_{-0.0130}$ | $0.3465^{+0.0250}_{-0.0352}$ | $77.92^{+2.64}_{-1.81}$ | R | 16 |
| Jan 30, 2017 | 2457784.7589 | ±16 | 3298.0 | 4 | 0.25 | $0.1415^{+0.0035}_{-0.0091}$ | $0.3195^{+0.0173}_{-0.0205}$ | $79.93^{+2.45}_{-1.49}$ | I | 16 |
| Feb 07, 2017 | 2457792.6475 | ±66 | 3308.0 | 23 | 0.35 | $0.1322^{+0.0094}_{-0.0185}$ | $0.3841^{+0.0336}_{-0.0430}$ | $74.51^{+3.79}_{-2.27}$ | R | 16 |
| Feb 11, 2017 | 2457796.5914 | ±23 | 3313.0 | -5 | -0.22 | $0.1295^{+0.0044}_{-0.0084}$ | $0.2977^{+0.0293}_{-0.0241}$ | $80.68^{+3.09}_{-2.36}$ | clear | 6 |
| Apr 12, 2017 | 2457856.5423 | ±39 | 3389.0 | -82 | -2.10 | $0.1475^{+0.0034}_{-0.0046}$ | $0.3254^{+0.0108}_{-0.0120}$ | $78.78^{+0.90}_{-0.76}$ | clear | 6 |
| May 08, 2017 | 2457882.5749 | ±13 | 3422.0 | -5 | -0.38 | $0.1447 \pm 0.0020$ | $0.3280 \pm 0.0085$ | $79.27 \pm 0.74$ | R | 15 |
| Jun 11, 2017 | 2457916.4951 | ±11 | 3465.0 | 7 | 0.64 | $0.1414 \pm 0.0011$ | $0.3195 \pm 0.0058$ | $79.68 \pm 0.44$ | R | 15 |
| Nov 21, 2017 | 2458079.7857 | ±18 | 3672.0 | 86 | 4.78 | $0.1563^{+0.0042}_{-0.0122}$ | $0.3459^{+0.0160}_{-0.0208}$ | $77.73^{+2.03}_{-1.54}$ | R | 16 |
| Apr 04, 2018 | 2458213.8875 | ±18 | 3842.0 | 13 | 0.72 | $0.1393^{+0.0024}_{-0.0066}$ | $0.3118^{+0.0376}_{-0.0323}$ | $79.54^{+3.06}_{-3.00}$ | R | 6 |
| Apr 28, 2018 | 2458237.5526 | ±17 | 3872.0 | 12 | 0.71 | $0.1445 \pm 0.0034$ | $0.3276 \pm 0.0169$ | $79.24 \pm 1.74$ | I | 15 |
| May 02, 2018 | 2458241.4962 | ±13 | 3877.0 | -41 | -3.15 | $0.1427 \pm 0.0021$ | $0.3153 \pm 0.0086$ | $80.11 \pm 0.81$ | I | 15 |
| May 05, 2018 | 2458244.6522 | ±14 | 3881.0 | 15 | 1.07 | $0.1439 \pm 0.0022$ | $0.3276 \pm 0.0108$ | $79.00 \pm 1.02$ | I | 15 |
| Jun 27, 2018 | 2458282.5163 | ±11 | 3929.0 | -2 | -0.18 | $0.1392 \pm 0.0023$ | $0.3163 \pm 0.0088$ | $80.37 \pm 0.83$ | I | 15 |
| Feb 12, 2019 | 2458527.8458 | ±39 | 4240.0 | 42 | 1.08 | $0.1664^{+0.0186}_{-0.0141}$ | $0.3416^{+0.0300}_{-0.0357}$ | $77.97^{+1.81}_{-1.59}$ | R | 16 |
| Mar 06, 2019 | 2458549.1447 | ±41 | 4267.0 | 68 | 1.66 | $0.1311^{+0.0076}_{-0.0056}$ | $0.2712^{+0.0651}_{-0.0235}$ | $83.64^{+6.26}_{-3.93}$ | V | 6 |
| Mar 20, 2019 | 2458562.9490 | ±10 | 4284.5 | 37 | 3.70 | $0.1418 \pm 0.00536$ | $0.3220 \pm 0.0323$ | $80.76 \pm 3.29$ | R | 17 |
| Apr 07, 2019 | 2458580.6972 | ±35 | 4307.0 | -22 | -0.63 | $0.1410 \pm 0.015$ | $0.3215 \pm 0.017$ | $81.23 \pm 2.12$ | $I + z'$ | 13 |
| Apr 10, 2019 | 2458584.6417 | ±23 | 4312.0 | -2 | -0.09 | $0.1436^{+0.0021}_{-0.0029}$ | $0.3117^{+0.0115}_{-0.0113}$ | $79.84^{+1.07}_{-0.93}$ | R | 16 |
| May 10, 2019 | 2458614.6172 | ±14 | 4350.0 | -36 | -2.57 | $0.1446 \pm 0.0021$ | $0.3276 \pm 0.0095$ | $79.05 \pm 0.87$ | I | 15 |
| May 14, 2019 | 2458618.5615 | ±12 | 4355.0 | -27 | -2.25 | $0.1434 \pm 0.0013$ | $0.3261 \pm 0.0072$ | $79.15 \pm 0.64$ | I | 15 |
| Jun 02, 2019 | 2458637.4940 | ±12 | 4379.0 | 4 | -0.33 | $0.1444 \pm 0.0016$ | $0.3257 \pm 0.0070$ | $79.12 \pm 0.68$ | I | 15 |
| Mar 19, 2021 | 2459292.2304 | ±10 | 5210.0 | 2 | 0.20 | $0.1285 \pm 0.0070$ | $0.2566 \pm 0.0429$ | $88.59 \pm 4.34$ | R | 17 |
| May 23, 2021 | 2459358.4925 | ±11 | 5293.0 | -23 | -2.09 | $0.1447 \pm 0.0018$ | $0.3250 \pm 0.0081$ | $79.35 \pm 0.73$ | I | 15 |
| Jun 18, 2021 | 2459384.5245 | ±18 | 5326.0 | 1 | -0.06 | $0.1394 \pm 0.0017$ | $0.3127 \pm 0.0125$ | $79.71 \pm 1.08$ | I | 15 |
| Jun 08, 2022 | 2459739.5018 | ±16 | 5776.0 | -23 | -1.44 | $0.1436 \pm 0.0023$ | $0.3457 \pm 0.0112$ | $78.14 \pm 0.88$ | I | 15 |
| Feb 26, 2023 | 2460002.1858 | ±17 | 6109.0 | 27 | 1.59 | $0.1322 \pm 0.0051$ | $0.2685 \pm 0.0346$ | $85.09 \pm 3.77$ | R | 17 |
| Mar 23, 2023 | 2460026.6393 | ±14 | 6140.5 | -21 | -1.50 | $0.1452 \pm 0.0046$ | $0.3296 \pm 0.0249$ | $79.28 \pm 2.22$ | R | 17 |
| May 21, 2023 | 2460086.5911 | ±18 | 6216.0 | -18 | -1.00 | $0.1320 \pm 0.0041$ | $0.2787 \pm 0.0183$ | $83.53 \pm 2.11$ | I | 15 |
| Jun 05, 2023 | 2460101.5791 | ±16 | 6235.0 | -9 | -0.56 | $0.1462 \pm 0.0009$ | $0.3479 \pm 0.0077$ | $77.24 \pm 0.61$ | I | 15 |
| Mar 01, 2025 | 2460734.2280 | ±40 | 7037.0 | -13 | -0.33 | $0.1461 \pm 0.0063$ | $0.3286 \pm 0.033$ | $78.88 \pm 3.37$ | R | 17 |
| Mar 29, 2025 | 2460762.6258 | ±35 | 7073.0 | -47 | -1.34 | $0.1484 \pm 0.0055$ | $0.3484 \pm 0.0358$ | $77.96 \pm 2.96$ | R | 17 |

**Notes.** References: (1) Hebb et al. (2010); (2) Albrecht et al. (2012); (3) Anderson et al. (2013); (4) Tregloan-Reed et al. (2013); (5) Hellier et al. (2011); (6) Exoplanet Transit Database (ETD); (7) Mancini et al. (2013); (8) Lendl et al. (2013); (9) Dragomir et al. (2011); (10) Bean et al. (2013); (11) Espinoza et al. (2019); (12) Sedaghati et al. (2017); (13) Patra et al. (2020); (14) Sedaghati et al. (2015b); (15) This work (Danish telescope); (16) Petrucci et al. (2020); (17) This work (TESS)

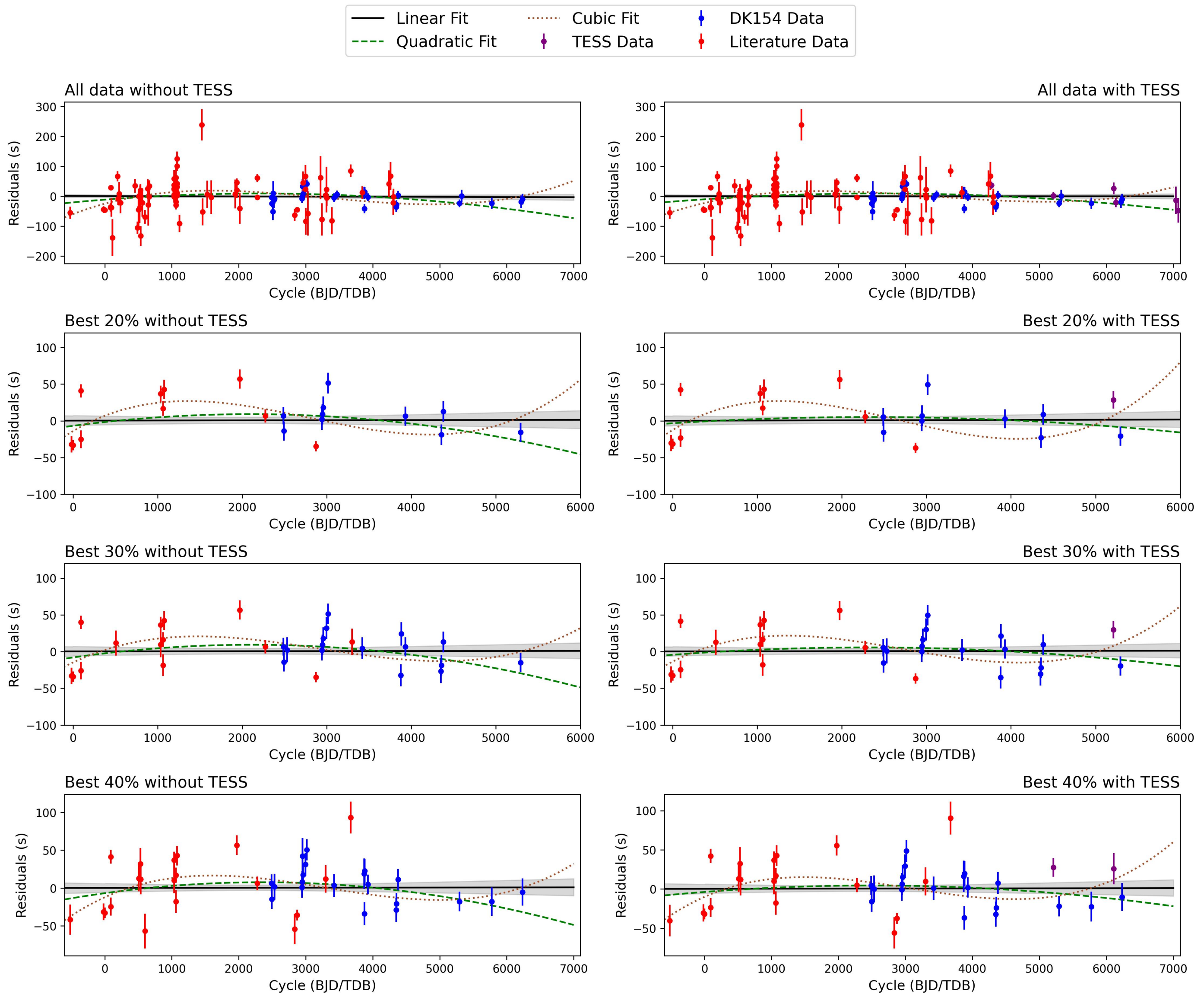

**Fig. A.2.** Transit timing residuals (O–C diagram) for WASP-19 b under different data inclusion scenarios. Each panel compares linear (solid black), quadratic (dashed green), and cubic (dotted brown) ephemeris models fitted to subsets of the data, with and without the inclusion of TESS transit timings. The top row shows results using all data, while the middle and bottom rows show fits restricted to the best 20%, 30%, and 40% of light curves based on photometric precision. Literature data are shown in red, Danish light curves (DK154) in blue, and TESS data in purple.

first calculating the $n^{th}$ transit time and then subtracting the first two terms representing linear ephemerides (Hadden & Lithwick 2016, 2017):

$$t(n) = T_0 + nP_b + \mu' \delta t^{(0)}(n) + \mu' \mathrm{Re}[Z] \delta t^{(1,x)}(n) + \mu' \mathrm{Im}[Z] \delta t^{(1,y)}(n) \,, \tag{B.1}$$

where $\mu'$ represents the perturber's planet-star mass ratio, and $t^{(0)}(n)$, $t^{(1,x)}(n)$, and $t^{(1,y)}(n)$ depend on the orbital period $P_b$ and the reference time (or transit epoch) $T_0$ of the inner planet. The variable $Z$:

$$Z = \frac{(z' - z)}{\sqrt{2}} \,, \tag{B.2}$$

contains the free complex eccentricities of the inner planet $z = ee^{i\overline{\omega}}$ and the outer planet $z' = e'e^{i\overline{\omega}'}$ (Hadden & Lithwick 2016, 2017), $\overline{\omega}$ and $\overline{\omega}'$ are their corresponding argument of periastron.

To derive the outer planet's TTV series, we simply need to switch the following quantities: $\mu' \rightarrow \mu$, $T_0 \rightarrow T_0'$, and $P_b \rightarrow P'$.

By fitting the measured TTVs with the synthetic TTV series, we can perform a separate MCMC to constrain the planetary mass ($M_p$) while exploring the parameter space of all input orbital parameters for `TTVFast`, including orbital period ($P_b$), eccentricity ($e$), inclination ($i$), longitude of the ascending node ($\Omega$), argument of periastron ($\overline{\omega}$), and mean anomaly ($M$). The constrained planetary mass is $M_p = 1.171^{+0.125}_{-0.111}\, M_{\mathrm{Jup}}$, which is consistent with the result reported in Table 2.

Furthermore, we extracted periodicity information from long-term synthetic TTV series using the Lomb-Scargle (LS) periodogram, a unique technique to detect and characterize periodic signals from unequally spaced data (Lomb 1976; Scargle 1982; VanderPlas 2018). In this paper, we applied this technique to explore the periodicity from the `TTVFast` model in Fig. 4 based on N-body simulations of gravitational interactions in planetary systems, not the cubic model in Fig. 3 that describes

**Table A.3.** Statistical comparison of linear, quadratic, and cubic ephemeris models for WASP-19 b under different data-quality selections.

| Probability | Without TESS | | | | With TESS | | | |
|---|---|---|---|---|---|---|---|---|
| | Quantity | Linear | Quadratic | Cubic | Quantity | Linear | Quadratic | Cubic |
| **All data** | $\chi^2$ | 532.46 | 493.65 | 433.72 | $\chi^2$ | 547.02 | 516.29 | 468.47 |
| | $\chi^2_\nu$ | 4.98 | 4.66 | 4.13 | $\chi^2_\nu$ | 4.84 | 4.61 | 4.22 |
| | AIC | 536.46 | 499.65 | 441.72 | AIC | 551.02 | 522.29 | 476.47 |
| | BIC | 541.84 | 507.73 | 452.48 | BIC | 556.51 | 530.53 | 487.45 |
| | rms (s) | 48.40 | 47.85 | 45.72 | rms (s) | 47.40 | 46.50 | 45.45 |
| | $X_{\rm MAD}$ | $3.25\times10^{-4}$ | $2.91\times10^{-4}$ | $2.51\times10^{-4}$ | $X_{\rm MAD}$ | $3.14\times10^{-4}$ | $2.90\times10^{-4}$ | $2.49\times10^{-4}$ |
| **Best 20%** | $\chi^2$ | 146.50 | 134.37 | 111.78 | $\chi^2$ | 151.42 | 148.51 | 114.08 |
| | $\chi^2_\nu$ | 8.14 | 7.90 | 6.99 | $\chi^2_\nu$ | 8.41 | 8.74 | 7.13 |
| | AIC | 150.50 | 140.37 | 119.78 | AIC | 155.42 | 154.51 | 122.08 |
| | BIC | 152.49 | 143.35 | 123.76 | BIC | 157.41 | 157.50 | 126.06 |
| | rms (s) | 28.73 | 26.67 | 25.45 | rms (s) | 29.12 | 28.42 | 26.11 |
| | $X_{\rm MAD}$ | $3.05\times10^{-4}$ | $2.80\times10^{-4}$ | $2.51\times10^{-4}$ | $X_{\rm MAD}$ | $3.54\times10^{-4}$ | $3.41\times10^{-4}$ | $2.59\times10^{-4}$ |
| **Best 30%** | $\chi^2$ | 163.82 | 148.72 | 133.44 | $\chi^2$ | 170.07 | 165.55 | 138.04 |
| | $\chi^2_\nu$ | 5.85 | 5.51 | 5.13 | $\chi^2_\nu$ | 6.07 | 6.13 | 5.31 |
| | AIC | 167.82 | 154.72 | 141.44 | AIC | 174.07 | 171.55 | 146.04 |
| | BIC | 170.62 | 158.93 | 147.04 | BIC | 176.87 | 175.75 | 151.65 |
| | rms (s) | 26.19 | 24.21 | 23.89 | rms (s) | 26.65 | 25.88 | 24.62 |
| | $X_{\rm MAD}$ | $2.42\times10^{-4}$ | $2.48\times10^{-4}$ | $2.26\times10^{-4}$ | $X_{\rm MAD}$ | $2.76\times10^{-4}$ | $2.62\times10^{-4}$ | $2.50\times10^{-4}$ |
| **Best 40%** | $\chi^2$ | 209.07 | 195.98 | 181.31 | $\chi^2$ | 207.89 | 203.88 | 184.86 |
| | $\chi^2_\nu$ | 5.50 | 5.30 | 5.04 | $\chi^2_\nu$ | 5.47 | 5.51 | 5.14 |
| | AIC | 213.07 | 201.98 | 189.31 | AIC | 211.89 | 209.88 | 192.86 |
| | BIC | 216.45 | 207.05 | 196.06 | BIC | 215.27 | 214.95 | 199.61 |
| | rms (s) | 31.96 | 30.61 | 30.36 | rms (s) | 30.44 | 29.98 | 29.12 |
| | $X_{\rm MAD}$ | $2.88\times10^{-4}$ | $2.47\times10^{-4}$ | $2.16\times10^{-4}$ | $X_{\rm MAD}$ | $2.80\times10^{-4}$ | $2.69\times10^{-4}$ | $2.53\times10^{-4}$ |

**Notes.** Results are shown with and without the inclusion of TESS sectors and are derived from bootstrap-resampled MCMC fits with $N_{\rm boot}$ = 2000. The reported $\chi^2$, reduced $\chi^2_\nu$, AIC, and BIC values are computed directly from the timing uncertainties listed in Table A.2, without additional uncertainty rescaling (see Sect. 5.1). The cross-validation metric $X_{\rm MAD}$ is obtained from random hold-out tests removing 20 % of the data and refitting the remaining 80 % over 10 000 iterations. Although none of the models achieves $\chi^2_\nu \approx 1$, the cubic ephemeris consistently yields lower information-criterion values and stable predictive performance across different subsets. The overall timing scatter limits the statistical significance of any single ephemeris model.

**Table A.4.** Statistical results of ephemeris model fits under different data-quality filtering criteria.

| Data selection criteria | Model | $\chi^2$ | $\chi^2_\nu$ | AIC | BIC |
|---|---|---|---|---|---|
| After removing 5% highest uncertainties | Linear | 540.55 | 5.005 | 544.55 | 549.95 |
| | Quadratic | 510.87 | 4.774 | 516.87 | 524.97 |
| | Cubic | 463.11 | 4.369 | 471.11 | 481.92 |
| After removing 1$\sigma$ residual outliers | Linear | 306.77 | 3.371 | 310.77 | 315.84 |
| | Quadratic | 282.41 | 3.138 | 288.41 | 296.01 |
| | Cubic | 242.23 | 2.722 | 250.23 | 260.36 |
| After removing 2$\sigma$ residual outliers | Linear | 412.27 | 3.853 | 416.27 | 421.65 |
| | Quadratic | 388.30 | 3.663 | 394.30 | 402.37 |
| | Cubic | 353.61 | 3.368 | 361.61 | 372.37 |

**Notes.** The table reports the total $\chi^2$, reduced $\chi^2_\nu$, Akaike Information Criterion (AIC), and Bayesian Information Criterion (BIC) for the full dataset, after removing the 5% most uncertain transit times, and after excluding residual outliers beyond the 1$\sigma$ and 2$\sigma$ thresholds. All statistics are computed consistently following the uncertainty treatment described in Sect. 5.1, allowing direct comparison of model performance across the different filtering tests.

slow, non-periodic curvature expected from apsidal precession or tidal dissipation effects. Similar methodology can be found in other papers, such as Holczer et al. (2016) and Yahalomi & Kipping (2026). The LS periodogram for WASP-19 b when searching for periodic signals up to $10^5$ d is illustrated in Fig. B.1, where WASP-19 b's orbital period and its sub-harmonics (in the form of $P_{\rm WASP-19\,b}/n$, $n = \{2, 3, 4, 5, 6, 7, 8, 9, 10\}$) are reflected as peaks in the upper panel. After removing all sub-harmonics in the middle panel of Fig. B.1, we found a power "excess" region at $P \gtrsim 1200$ d without long-period peaks being detected. The false alarm probability (FAP) returns to zero in this region, indicating that the LS signal is statistically strong and confident. Although the middle panel of Fig. B.1 shows a broad “excess” of power at periods $P \gtrsim 1200$ d once the sub-harmonics of the planetary orbital period are removed, this excess does not manifest as a distinct long-period peak. To assess whether this feature could still correspond to a genuine periodic modulation, we computed both analytic and bootstrap false-alarm probabilities (FAPs) using the measured TTV residuals.

The analytic FAP was evaluated using the classical expression for the exponential tail of the Lomb–Scargle periodogram under the null hypothesis of Gaussian noise (Lomb 1976; Scar-

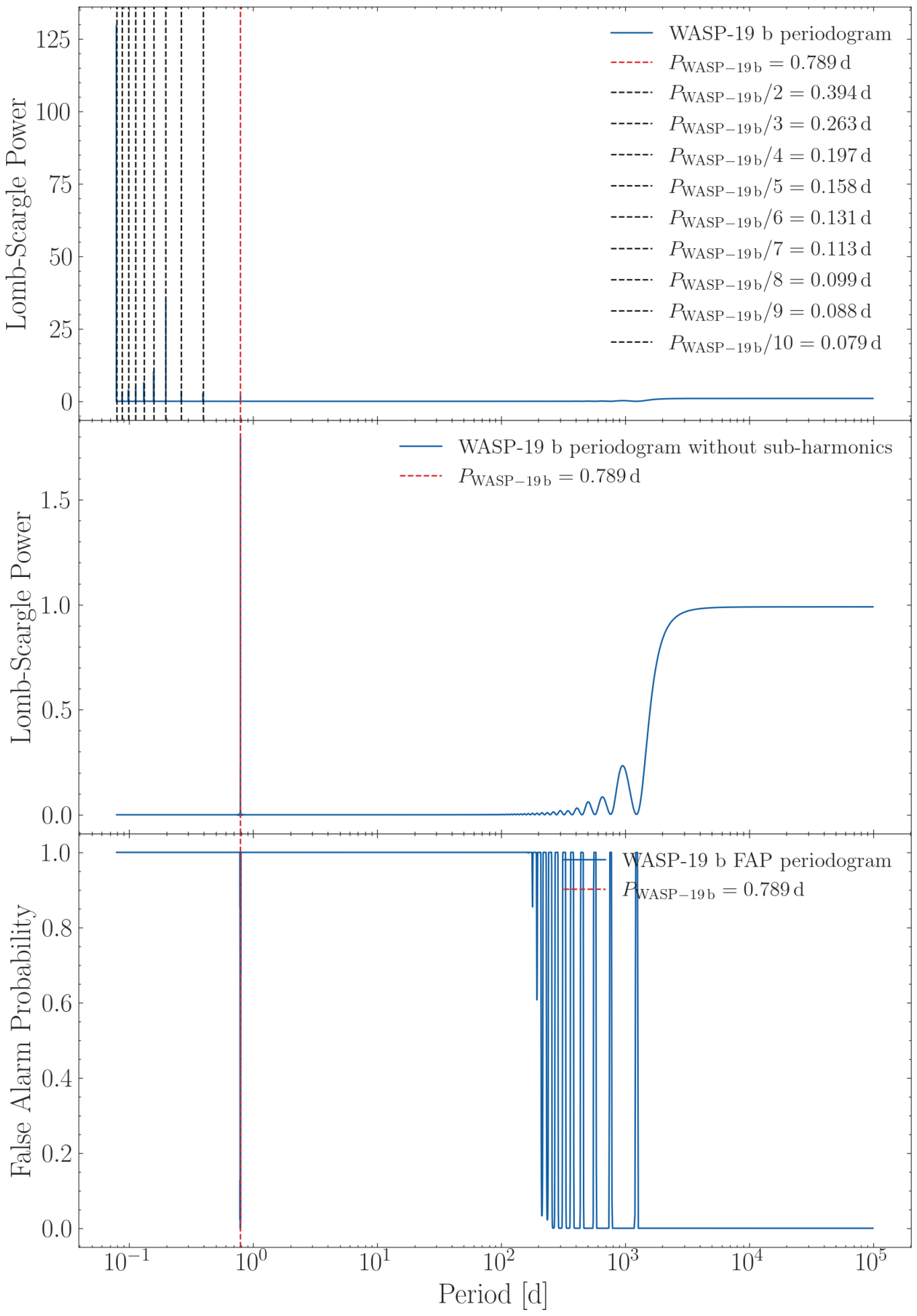

**Fig. B.1.** LS periodogram for WASP-19 b. The black dashed lines in the upper panel represent sub-harmonics of WASP-19 b's periodic signal at $P \approx 0.789$ d labelled with red dashed lines, while all sub-harmonics are removed in the middle panel. As a contrast, the corresponding FAP periodogram is illustrated in the lower panel. A long-period power "excess" region at $P \gtrsim 1200$ d is observed in the LS periodogram, but no obvious distant peaks are detected.

gle 1982; VanderPlas 2018):

$$\mathrm{FAP}_{\rm analytic} = 1 - [1 - e^{-z}]^{N_{\rm eff}}\,,$$

where $z$ is the normalized LS power and $N_{\rm eff}$ the number of independent frequencies. Applying this to the global peak power ($z = 0.118$) yields $\simeq 0.95$, indicating that the observed long-period excess is fully consistent with noise fluctuations.

To account for irregular sampling, we also performed a non-parametric bootstrap test $\mathrm{FAP}_{\rm bootstrap}$ by randomly shuffling the TTV residuals while keeping observation epochs fixed. For each of $10^4$ realizations, a Lomb–Scargle periodogram was recomputed, and the highest peak was recorded. The resulting empirical false-alarm probability yields $\simeq 0.981$ showing that nearly all bootstrap realizations produce peaks equal to or greater than the observed excess.

Together, the analytic and bootstrap tests demonstrate that the long-period power excess arises from baseline-limited spectral leakage rather than a coherent dynamical signal. This confirms that the LS periodogram does not contain statistically significant evidence of a long-period TTV modulation, and therefore does not support the presence of an additional planetary perturber in the WASP-19 system.

## Appendix C: Baseline correction for transmission photometric spectrum

To construct a consistent transmission spectrum from heterogeneous datasets, we applied a constant additive offset in transit depth to each instrument's measurements. We adopted the HST STIS+WFC3 spectrum (Huitson et al. 2013) as the reference baseline due to its high signal-to-noise ratio, stable systematics, and continuous optical–near-IR coverage. For each ground-based or single-band dataset, we interpolated the HST spectrum onto the wavelength grid of that dataset within the overlapping region and solved for a constant offset $\Delta d$ that minimized the weighted squared difference between the datasets. Uncertainties in $\Delta d$ were estimated via bootstrap resampling. The resulting offsets are physically small (typically $|\Delta d| \lesssim 1.2 \times 10^{-3}$), consistent with expected variations from unocculted stellar spots and instrumental baseline differences: for example, Huitson et al. (2013) report that unocculted spots can shift $R_{\rm p}/R_\star$ by up to 1.38% between epochs, corresponding to $\Delta d \sim (0.4–8) \times 10^{-4}$, while Sedaghati et al. (2015a) estimate a baseline uncertainty of $\sim 0.0017$ in $R_{\rm p}/R_\star$ from FORS2 systematics. Similar vertical offsets are commonly applied in multi-instrument analyses (e.g., Bean et al. 2013; Mancini et al. 2013). Although we explored a broad prior range of $\Delta d \in [-0.02, +0.02]$ in transit depth, the best-fit values are more than an order of magnitude smaller, and for subsequent retrievals a tighter prior (e.g., $\Delta d \in [-5 \times 10^{-3}, +5 \times 10^{-3}]$ or a Gaussian prior with $\sigma \simeq 0.001$) is physically motivated. The final offsets applied in this work are shown in Table C. This homogenization ensures that the combined spectrum preserves the true wavelength-dependent atmospheric signal without being biased by instrument-specific baselines.

**Table C.1.** Offset corrections applied to each WASP-19 b dataset.

| Dataset / Instrument | Offset (transit depth) | $1\sigma$ | $N_{\rm overlap}$ |
|---|---|---|---|
| **(Huitson et al. 2013)** | +0.000000 | 0.000000 | 10 |
| (Bean et al. 2013) | **−0.000966** | 0.000036 | 3 |
| (Mancini et al. 2013) | **−0.000653** | 0.000103 | 4 |
| (Sedaghati et al. 2015b) | **−0.000422** | 0.000093 | 25 |
| Danish (This work) | **−0.001202** | 0.000256 | 2 |
| TESS (This work) | **−0.000397** | 0.000329 | 1 |

**Notes.** Offsets are computed relative to the combined HST STIS+WFC3 transmission spectrum and applied uniformly to align the datasets.

**Table C.2.** Derived photometric parameters from each light curve, plus the interval within which the best fit was searched for using GEMC.

| Parameter | Symbol | Search Interval | 24/02/2010 | 25/02/2010 | 28/02/2010 | 05/06/2010 | 20/06/2010 | 11/05/2011 |
|---|---|---|---|---|---|---|---|---|
| | | | Tregloan-Reed et al. (2013) | Tregloan-Reed et al. (2013) | Tregloan-Reed et al. (2013) | Mancini et al. (2013) | Mancini et al. (2013) | Mancini et al. (2013) |
| Radius ratio | $r_p/r_s$ | 0.1 to 0.2 | 0.1406±0.0020 | 0.1369±0.0015 | 0.1446±0.0019 | 0.1390±0.0023 | 0.1478±0.0034 | 0.1376±0.0011 |
| Sum of fractional radii | $r_p + r_s$ | 0.3 to 0.4 | 0.3205±0.0084 | 0.3132±0.0056 | 0.3361±0.0070 | 0.3128±0.0117 | 0.3442±0.0170 | 0.3276±0.0082 |
| Linear LD coefficient | $u_1$ | 0.4 to 0.6 | 0.174±0.127 | 0.499±0.132 | 0.548±0.135 | 0.096±0.157 | 0.273±0.186 | 0.268±0.127 |
| Quadratic LD coefficient | $u_2$ | 0.2 to 0.4 | 0.2753±0.1843 | 0.352±0.180 | 0.016±0.213 | 0.621±0.237 | 0.351±0.240 | 0.359±0.169 |
| Incination (degrees) | $i$ | 78 to 80 | 79.53±0.75 | 80.52±0.46 | 78.50±0.66 | 80.82±1.12 | 79.00±1.43 | 79.26±0.65 |
| Transit epoch (BJD/TDB) | $T_0$ | ±0.5 in phase | 2455251.795898229 ±0.000158656 | 2455252.585107156 ±0.000129384 | 2455255.740383137 ±0.000125570 | 2455353.951391019 ±0.048889074 | 2455368.544568784 ±0.000350326 | 2455693.940599460 ±0.034687415 |
| Longitude of spot (degrees) | $\theta$ | -90 to +90 | -10.78±2.68 | 14.72±2.02 | | 10.15±2.55 | 29.78±14.45 | -12.69±25.31 |
| Latitude of spot (degrees) | $\phi$ | 0.0 to 90.0 | 18.46±10.06 | 21.17±4.91 | | 22.71±27.47 | 46.40±14.34 | 27.70±13.59 |
| Spot angular radius (degrees) | $r_{spot}$ | 0.0 to 30.0 | 17.77±6.50 | 17.54±3.79 | | 7.01±14.13 | 12.43±15.17 | 11.30±12.80 |
| Spot contrast | $\rho_{spot}$ | 0.5 to 1.0 | 0.68±0.07 | 0.85±0.06 | | 0.25±0.12 | 0.83±0.23 | 0.77±0.18 |

| Parameter | Symbol | Search interval | 15/04/2012 | 15/04/2012 | 15/04/2012 | 15/04/2012 | 15/04/2012 | 15/04/2012 |
|---|---|---|---|---|---|---|---|---|
| | | | Mancini et al. (2013) | Mancini et al. (2013) | Mancini et al. (2013) | Mancini et al. (2013) | Mancini et al. (2013) | Mancini et al. (2013) |
| Radius ratio | $r_p/r_s$ | 0.1 to 0.2 | 0.1417±0.0019 | 0.1434±0.0015 | 0.1417±0.0020 | 0.1414±0.0014 | 0.1310±0.0018 | 0.1405±0.0041 |
| Sum of fractional radii | $r_p + r_s$ | 0.3 to 0.4 | 0.3291±0.0082 | 0.3318±0.0072 | 0.3262±0.0079 | 0.3333±0.0066 | 0.3239±0.0095 | 0.3439±0.0190 |
| Linear LD coefficient | $u_1$ | 0.4 to 0.6 | 0.447±0.172 | 0.280±0.158 | 0.094±0.172 | 0.220±0.131 | 0.166±0.100 | 0.472±0.233 |
| Quadratic LD coefficient | $u_2$ | 0.2 to 0.4 | 0.096±0.225 | 0.222±0.201 | 0.487±0.247 | 0.218±0.143 | 0.032±0.149 | 0.448±0.246 |
| Incination (degrees) | $i$ | 78 to 80 | 78.63±0.71 | 78.64±0.66 | 79.12±0.76 | 78.70±0.54 | 78.96±0.82 | 78.57±1.58 |
| Transit epoch (BJD/TDB) | $T_0$ | ±0.5 in phase | 2456033.536172090 ±0.000143092 | 2456033.536276974 ±0.000140267 | 2456033.536253734 ±0.000138668 | 2456033.536513193 ±0.000131707 | 2456033.537318070 ±0.000160769 | 2456033.537691631 ±0.000231036 |
| Longitude of spot (degrees) | $\theta$ | -90 to +90 | 10.02±15.12 | 10.02±15.12 | 10.02±15.12 | 10.02±15.12 | 10.02±15.12 | 10.02±15.12 |
| Latitude of spot (degrees) | $\phi$ | 0.0 to 90.0 | 28.75±5.30 | 31.60±5.25 | 36.27±5.39 | 37.12±5.39 | 35.99±11.29 | 34.73±12.82 |
| Spot angular radius (degrees) | $r_{spot}$ | 0.0 to 30.0 | 9.33±3.26 | 6.82±3.33 | 5.85±3.49 | 3.60±6.28 | 4.34±8.72 | 10.87±9.28 |
| Spot contrast | $\rho_{spot}$ | 0.5 to 1.0 | 0.12±0.19 | 0.27±0.16 | 0.66±0.13 | 0.44±0.15 | 0.56±0.24 | 0.83±0.22 |

| Parameter | Symbol | Search Interval | 15/04/2012 | 15/04/2012 | 03/05/2015 | 07/05/2015 | 18/05/2015 | 11/06/2017 |
|---|---|---|---|---|---|---|---|---|
| | | | Mancini et al. (2013) | Lendl et al. (2013) | Danish (this work) | Danish (this work) | Danish (this work) | Danish (this work) |
| Radius ratio | $r_p/r_s$ | 0.1 to 0.2 | 0.1296±0.0033 | 0.1458±0.0023 | 0.1437±0.0011 | 0.1404±0.0017 | 0.1440±0.0013 | 0.1414±0.0011 |
| Sum of fractional radii | $r_p + r_s$ | 0.3 to 0.4 | 0.3071±0.0169 | 0.3438±0.0116 | 0.3210±0.0056 | 0.3277±0.0070 | 0.3323±0.0042 | 0.3195±0.0058 |
| Linear LD coefficient | $u_1$ | 0.4 to 0.6 | 0.016±0.127 | 0.325±0.138 | 0.496±0.110 | 0.754±0.128 | 0.5103±0.0276 | 0.320±0.129 |
| Quadratic LD coefficient | $u_2$ | 0.2 to 0.4 | 0.244±0.207 | 0.032±0.200 | 0.292±0.128 | 0.062±0.165 | 0.231±0.021 | 0.316±0.143 |
| Inclination (degrees) | $i$ | 78 to 80 | 80.20±1.64 | 78.00±0.95 | 80.01±0.51 | 79.67±0.60 | 78.95±0.37 | 79.68±0.44 |
| Transit epoch (BJD/TDB) | $T_0$ | ±0.5 in phase | 2456033.536524120 ±0.000245909 | 245606.773444534 ±0.000223983 | 2457146.587234485 ±0.000119938 | 2457150.531188892 ±0.000010245 | 2457161.574626629 ±0.000287633 | 2457916.494274837 ±0.000130124 |
| Longitude of spot (degrees) | $\theta$ | -90 to +90 | 10.02±15.12 | 16.08±40.10 | -2.688±2.048 | -7.693±1.693 | 9.155±1.396 | 32.037±13.072 |
| Latitude of spot (degrees) | $\phi$ | 0.0 to 90.0 | 36.31±11.75 | 21.53±15.15 | 34.645±4.687 | 20.730±4.381 | 26.249±0.960 | 23.965±5.685 |
| Spot angular radius (degrees) | $r_{spot}$ | 0.0 to 30.0 | 5.68±8.37 | 18.46±12.57 | 10.11±4.27 | 13.14±3.80 | 13.81±1.68 | 13.373±5.262 |
| Spot contrast | $\rho_{spot}$ | 0.5 to 1.0 | 0.61±0.22 | 0.80±0.25 | 0.915±0.106 | 0.557±0.134 | 0.83±0.06 | 0.787±0.212 |